\documentclass[aps,prx,twocolumn,floatfix,longbibliography,nofootinbib]{revtex4-2}
\usepackage[utf8]{inputenc}
\usepackage{natbib}
\usepackage{graphicx}
\usepackage{xcolor}
\usepackage[tbtags]{amsmath}
\usepackage[colorlinks, linkcolor=blue]{hyperref}
\hypersetup{colorlinks,allcolors=black}
\usepackage{amssymb}
\usepackage{gensymb}
\usepackage{txfonts}
\usepackage{comment}
\usepackage{float}
\usepackage{amsmath}
\usepackage{tabularx,graphicx}
\usepackage{epstopdf}
\usepackage{latexsym}
\usepackage{color, colortbl}
\usepackage{psfrag}
\usepackage{bbm}
\usepackage{bm}
\usepackage{titlesec}
\usepackage{dsfont}
\usepackage{feynmp}
\usepackage{slashed}
\usepackage{multirow}
\usepackage[normalem]{ulem}
\renewcommand{\vec}[1]{\mathbf{#1}}

\def \k {{\vec k}}
\def \p {{\vec p}}

\def \ve {\varepsilon}
\def \r {{\vec r}}

\def \l {\ell}
\def \ve {\varepsilon}

\def \beq {\begin{eqnarray}}
\def \eeq {\end{eqnarray}}
\def \tn {\textnormal}

\def \nn {\nonumber}

\newcommand{\avg}[1]{\left\langle #1 \right\rangle}

\begin{document}
\title{Theory of Correlated Insulators and Superconductor at $\nu=1$ in Twisted WSe$_2$} 
\author{Sunghoon Kim}\thanks{These authors contributed equally to this work}
\author{Juan Felipe Mendez-Valderrama}\thanks{These authors contributed equally to this work}
\author{Xuepeng Wang}\thanks{These authors contributed equally to this work}
\author{Debanjan Chowdhury}\email{Corresponding author: debanjanchowdhury@cornell.edu}
\affiliation{Department of Physics, Cornell University, Ithaca, New York 14853, USA.}
\begin{abstract}
The observation of a superconducting phase, an intertwined insulating phase, and a continuous transition between the two at a commensurate filling of $\nu=1$ in bilayers of twisted WSe$_2$ at $\theta=3.65^0$ raises a number of intriguing questions about the origin of this phenomenology. {Here we report the possibility of a displacement-field induced continuous transition between a superconductor and a quantum spin-liquid Mott insulator at $\nu=1$, starting with a simplified three-orbital model of twisted WSe$_2$, including on-site, nearest-neighbor density-density interactions, and a chiral-exchange interaction, respectively}. By employing parton mean-field theory, we discuss the nature of these correlated insulators, their expected evolution with the displacement-field, and their phenomenological properties.  
\end{abstract}

\maketitle

\section{Introduction} Superconductivity (SC) in two-dimensional (2D) electronic materials at low carrier densities has captivated the attention of physicists in recent years. The observation of SC in moir\'e \cite{Cao2018b,Yankowitz_2019,Lu2019,Arora2020SC,Hao2021TTGelectric} as well as moir\'e-less graphene \cite{Oh2021unconventional,Zhou2021RTGSC,Zhou2022BBGSC,Zhang2023BBGSC,Zhang2023BBGSC} in the vicinity of correlation-induced insulators \cite{Cao2018} and spontaneously spin (or valley) polarized metallic states \cite{Zhou2021RTGhalf} has raised the question of the extent to which pairing is due to the proximate electronic orders. The role of electron-electron vs. electron-phonon interactions in inducing SC in these platforms has also been scrutinized intensely, even as the experimental situation remains largely unclear \cite{saito2020independent,Stepanov_2020,liu2021tuning}. The recent discovery of superconductivity and an intertwined correlated insulator in twisted bilayers of WSe$_2$ (tWSe$_2$) near $\theta=3.65^0$ only at a commensurate filling \cite{TMDSC} present a number of fascinating puzzles that requires a critical examination of strong-coupling effects, originating from electronic interactions. Superconductivity has also been reported at a larger twist-angle in twisted WSe$_2$ \cite{Dean24}, where the bare electronic bandwidth is higher and there are no proximate insulating phases in the phase-diagram, suggesting weaker effective correlations.

While previous experimental work \cite{Dean20} argued for possible signatures of SC in a doped insulator in tWSe$_2$, the recent report \cite{TMDSC} highlights a number of unconventional features tied to its origin. We highlight below some of the important phenomenological observations at the smaller twist-angle, which need to be taken into serious consideration from the outset and will be the focus of our attention here, and which point towards a strong-coupling perspective, beyond a purely fermiology-driven paradigm. First and foremost, the superconducting region occurs only in the vicinity of the commensurate filling $\nu=1$, and away from the van-Hove filling. Second, the predominant phase at $\nu=1$ is a correlated interaction-induced insulator, which only gives way to SC over a narrow range of displacement fields near $E_z=0$ below $T_c$. Notably, both the insulating and superconducting phases appear in the layer-hybridized regime, as opposed to the layer-polarized regime. Third, there appears to be a displacement field-induced continuous and direct superconductor-insulator transition at $\nu=1$. The insulator yields no topological response, in as far as electrical transport is concerned, and reveals fluctuating local-moments at high temperatures.

At first glance, the superconductor-insulator transition suggests the possibility of the insulator being a failed-superconductor \cite{SKRMP} --- a localized crystal of phase-incoherent (electronic) Cooper-pairs. However, the appearance of the local pairing only in the vicinity of $\nu=1$, rather than a wider range of dopings \cite{DCPRL23} suggests that the origin of pairing must be tied to the proximate (commensurate) Mott insulator. The experimental data suggests that the origin of pairing, or the glue, is potentially present in the insulator itself. In other words, it is not the pairing of electrons, but of other particles (e.g. spinons) in the parent insulating phase, that might be responsible for the subsidiary electronic pairing in the superconducting phase, separated from the parent insulator via a quantum phase transition. Our proposed scenario is distinct from a weak-coupling electron fermiology-driven instability \cite{constantin,SDS_VHS_RG,Senechal_VHS_dmft,Scherer_VHS_hetero,Kennes_VHS,HongYao_VHS,Biborski_VHS,Rademaker_VHS_strongcoupling} or a doping-induced instability \cite{EAKim_doped_dmrg,LiangFu_doped_trimer,DNSheng_doped_dmrg,Millis_doped_TSC,Yahui_doped_parton,KTLaw_doped}, which is generically not expected to yield a direct continuous superconductor-insulator transition at a fixed commensurate filling.

In this work, we analyze the above scenario for the interplay between the insulator and superconductor at the commensurate filling $\nu=1$. We start from a model Hamiltonian that is believed to capture many of the essential microscopic details of the electronic bandstructure, topological character, and interactions. Our basic proposal for the phenomenology at $\nu=1$ is  of a fully gapped quantum spin liquid insulator \cite{QSL,LNW}, where the electron fractionalizes into neutral fermionic spinons which are paired, and a gapped holon which carries the electric charge. As noted above, the fermionic pairing is arising within the insulator itself and the transition into the electronic superconductor at a fixed filling arises once the holons condense as a function of the displacement-field \cite{Senthil_Z2QSL_PRB}. Clearly, the doped quantum spin liquid can, in principle, harbor SC, as will also be demonstrated within the same parton mean-field computations. However, we discuss at the end on why this tendency can be suppressed in the present setting, and why this remains one of the exciting directions for theory in future work.

\begin{figure}[pth!]
\centering
\includegraphics[width=\linewidth]{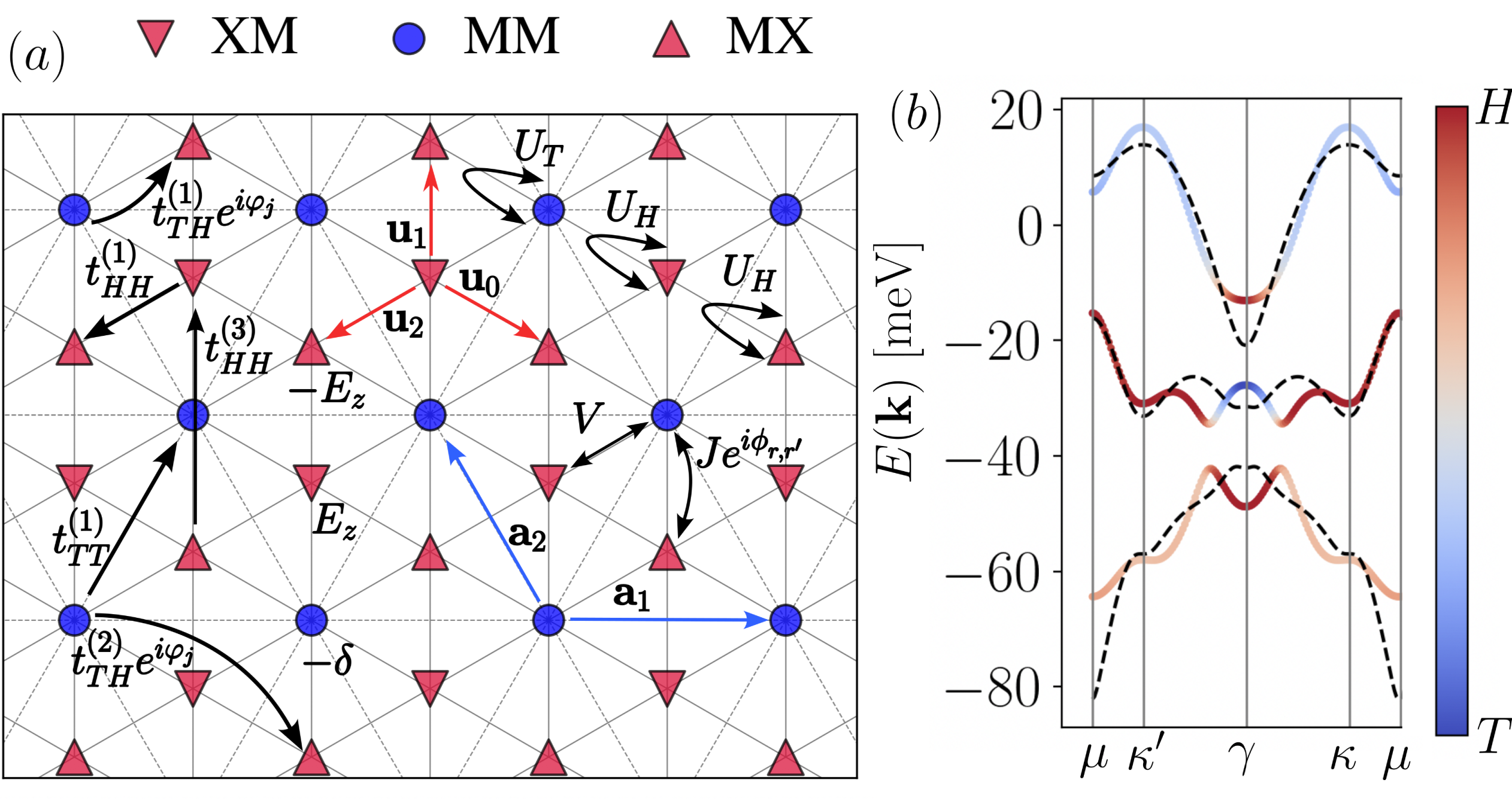} 
\caption{{\bf Three-orbital model for twisted WSe$_2$}. (a)  Schematic of the interacting three-orbital model for tWSe$_2$, where  $\bullet,\blacktriangledown,\blacktriangle$ symbols represent MM, XM and MX sites, respectively; see Eq.~\ref{Eq1:H0} and \ref{Eq1:Hint}. (b) Non-interacting electronic dispersions of the continuum model (dashed lines) and the tight-binding fit to the three-orbital model (solid colored line), respectively, for the topmost, second, and third moir\'e valence bands. The color coding denotes honeycomb (H) for the MX/XM sites, and triangular (T) for the MM site.} 
\label{fig:model}
\end{figure}

\section{Results}

\subsection{Model}
To demonstrate the possibility of the above scenario in a concrete setting, we start from a three-orbital electronic model \cite{valentin24} obtained from an underlying continuum model \cite{Wu_twisted_prl,Devakul_magic2021}. The general features of the model derive from taking the quadratic approximation for the topmost valence band of the monolayer valley, $K$, which is spin-split due to the strong spin-orbit coupling \cite{Xiao_SOC_PRL}. The opposite valley is related by time-reversal symmetry (TRS) and for AA stacking, the bands from both layers will feature spin $\uparrow (\downarrow)$ character for valley $K \left(-K\right)$. The bands for the bottom and top layer are slightly displaced to the corners of the Brillouin zone $\mathbf{\kappa}_{\pm}$, whose location is determined by the twist angle $\theta$. For WSe$_{2}$, the interlayer tunneling and the moir\'e potential has been determined from large-scale DFT calculations \cite{Devakul_magic2021}. At large twist angles, the two top-most bands in the continuum model feature equal valley contrasting Chern numbers. Consequently, to capture the low energy physics of these bands, a minimal model including at least three orbitals is needed to achieve a local real space description \cite{valentin24,fang_invariants_2012,3orb_Yu_NSR19,3orb_Wu_PRX24,3orb_Xu_PNAS24}.

We focus on the following interacting model in what follows \cite{valentin24}: $H=\sum_{\sigma=\uparrow,\downarrow} H_\sigma + H_{\tn{int}}$, where
\begin{subequations}
\beq   H_{\uparrow} &=& \sum_{\mathbf{k},\tau}c_{\mathbf{k},\tau}^{\dagger}\left[h_{1,\tau}\left(\mathbf{k}\right)+h_{2,\tau}\left(\mathbf{k}\right)\right]c_{\mathbf{k},\tau},\label{Eq1:H0}\\
H_{\tn{int}} &=& U_H\sum_{\r\in H} n_{\r\uparrow}n_{\r\downarrow} + U_T\sum_{\r\in T} n_{\r\uparrow}n_{\r\downarrow} + V \sum'_{\r\in T,\r'\in H} n_{\r}n_{\r'} \nn\\
&+& J \sum'_{\r\in T,\r'\in H} \bigg[e^{i\phi_{\r,\r'}} c^{\dagger}_{\r\uparrow}c^{\phantom\dagger}_{\r\downarrow}c^{\dagger}_{\r' \downarrow}c^{\phantom\dagger}_{\r' \uparrow} + \tn{h.c.} \bigg].\label{Eq1:Hint}
\eeq    
\end{subequations} 
Here, $c_{\mathbf{k},\tau}$ represents a spinor in orbital space, and $h_1(\mathbf{k}),~h_2(\mathbf{k})$ represent the nearest and further range hopping matrix elements, respectively. In the absence of the displacement field ($E_z=0$), the MX/XM sites are degenerate in energy, and split by a constant energy $\sim\delta$ relative to the energy of the MM sites; see Methods (Sec.~\ref{sec:Methods}) for details.

The choice of hoppings in the model is able to broadly capture the band topology of twisted TMD homobilayers by reproducing the $C_3$ eigenvalues of orbitals obtained from the continuum model at $\gamma,\kappa,$ and $\kappa'$, which can be done by retaining nearest neighbour hoppings in $h_1(\bold{k})$ and is consistent with DFT \cite{valentin24,Devakul_magic2021}. By additionally including further neighbour hoppings, the energetics of the topmost band can be reproduced to a high degree of accuracy at small twist angles. However, for $\theta=3.65^\circ$, the top three bands of the continuum model for twisted WSe$_2$ feature a combination of topological indices that disallows a local description \cite{Haining_band_topo_twse_2020}, which is reflected in the error incurred in the fit to the three orbital model  in Fig.\ref{fig:model}. Nevertheless, the uncertainties in the parametrization of the continuum model leads us to focus on capturing only the topology of the two topmost bands. Unlike experimental evidence for the non-trivial topology associated with the two topmost bands \cite{AustinExpt,Foutty_2024}, there is no such available evidence for the nontrivial topology tied to the third band at present. While the insulator at $\nu=1$ does not show any clear experimental indications of topological edge-states or orbital ferromagnetism, a more detailed future study can help clarify the role of band-topology in the correlation-driven Mott insulator. Phenomenologically, this allows us to study the low-energy physics of the experiment within a local description, but setting up the problem directly in momentum-space remains an interesting future direction. Even with this approximation, the Berry curvature distribution and quantum geometry of the topmost two bands can be reproduced faithfully from the continuum model. The integral of the Fubini-Study metric shows a particularly weak dependence on displacement field suggesting that localization of the Wannier orbitals at the level of the non-interacting bands is playing a subsidiary role and the main effect of the displacement field may be to introduce a sublattice potential difference between XM and MX stacking regions. 

Turning now to the interactions (Eq.~\ref{Eq1:Hint}), we have included an on-site repulsion $U_H,~U_T$ on the honeycomb and triangular sites, respectively, in addition to a nearest-neighbor (represented by $\sum'$) repulsion, $V$. Finally, the chiral-exchange interaction, $J$, arises between the $T$ and $H$ sites directly by projecting the Coulomb-interactions to the relevant bands, and the phase-factors $\phi_{\r,r'}=2\pi n/3$ with $n$ an integer that increases conter-clockwise labeling the six nearest neighbours around a $T$ site. This term preserves the time-reversal symmetry as it can be rewritten as the weighted sum of a Heisenberg and a two-spin Dzyaloshinskii-Moriya interaction. Note that we have not included a super-exchange interaction across the two valleys in the above description since it is expected to be small; nevertheless, such an interaction will also drive the same pairing tendency of spinons \cite{yahui_chiral_2023}. In order to analyze the phases and phase-transitions associated with the above model, we employ the parton representation for the electronic operators ($c = b f$, with $b$ and $f$ being charged boson and neutral spinon, respectively); see Methods (Sec.~\ref{sec:Methods}) and Supplementary Materials Sec.~I for details. We will begin by incorporating the mean-field decomposition of $H_{\tn{int}}$, where the effect of the $U,~V-$ terms associated with the on-site and nearest-neighbor repulsions at the commensurate filling are included in the bosonic sector, and the effect of the chiral-exchange $J-$ term is included in the fermionic sector, respectively. As a result, the Mottness associated with the repulsive interactions affects the holons and is expected to drive a superfluid-Mott transition at a fixed commensurate filling. 

\begin{figure}[pth!]
\centering
\includegraphics[width=1.0\linewidth]{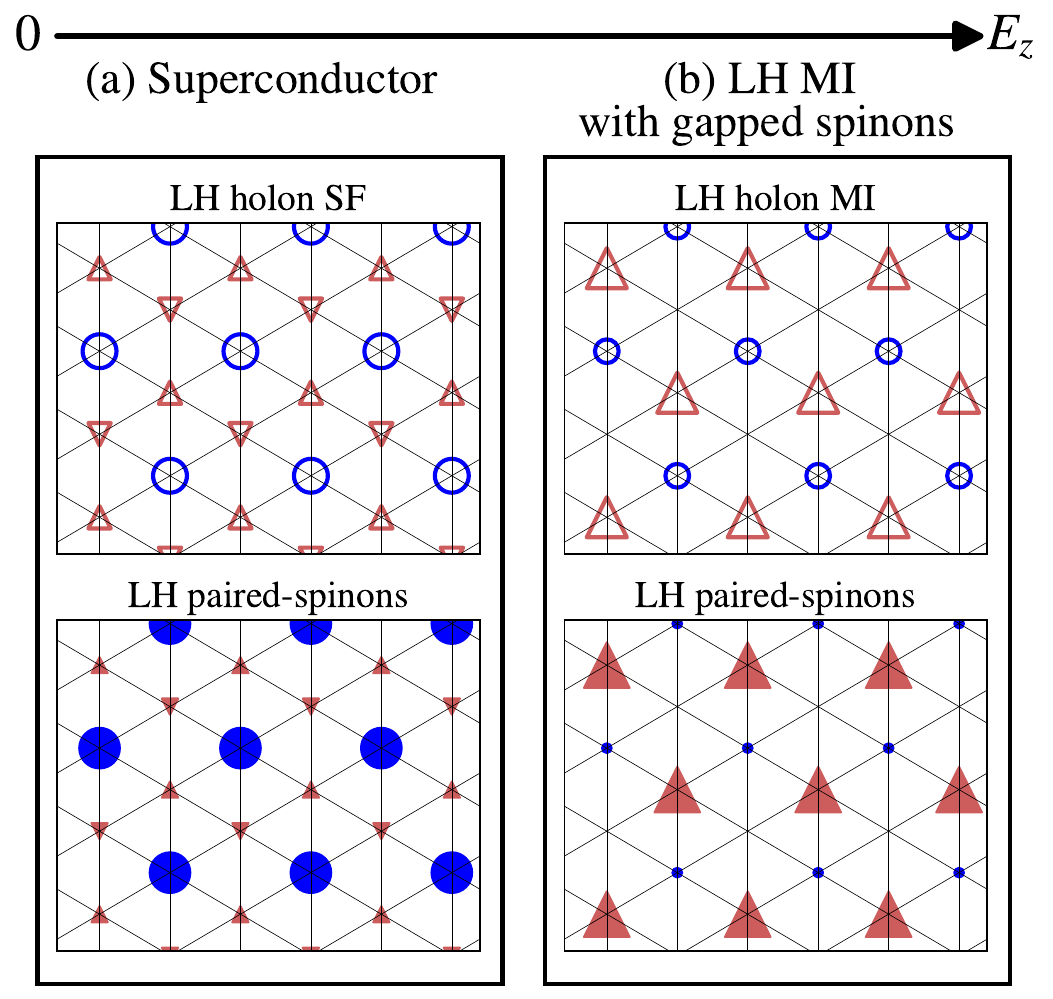} 
\caption{{\bf Evolution of the local holon and spinon occupations with increasing displacement-field}. The upper and lower panels show the occupations for the holons (upper panel) and spinons (lower panel) obtained from the parton mean-field computation for the Hamiltonian in Eq.~\ref{Eq1:H0} and Eq.~\ref{Eq1:Hint}. The empty ($\circ,\triangledown,\bigtriangleup$) and filled ($\bullet,\blacktriangledown,\blacktriangle$) symbols represent holon and spinon occupations, respectively, on the MM, XM and MX sites. The size of the markers is set so that the area of the bounding box surrounding them is proportional to the corresponding occupations determined from the parton mean-field computations. LH MI stands for layer-hybridized Mott insulator. (a) For small $E_z$, both the holons and spinons are LH; the spinons pair leading to an electronic superconductor when the holons are condensed (see text). (b) With increasing $E_z$, the holons are localized at MM and MX sites, forming an ``excitonic" Mott insulator, and thus retain their LH character (see text). The spinons are also LH and remain paired, resulting in an electronic MI with charge and spin-gaps, respectively. The plots in (a) and (b) are generated for values of the electric field corresponding to the red and green $\times$-markers in Fig.~\ref{fig3_boson} and Fig.~\ref{fig4_spinon}, respectively. With large $E_z$, the spinons and holons localize on the MX sites in a layer-polarized Mott insulator; see the supplementary material Sec.~\ref{sec:SI_LPMI}.}
\label{fig:occupation}
\end{figure}

\subsection{Field-induced superconductor-insulator transition}
Let us begin by describing the patterns of the holon and spinon (de-)localization in real-space on the different sublattice sites based on the solution of the parton mean-field equations (see Methods in Sec.~\ref{sec:Methods}), and summarizing the essential theoretical results. As shown in Fig.~\ref{fig:occupation}, the relative marker sizes are obtained from the solutions to the fully self-consistent parton mean-field computations, accounting for a superconducting and insulating spin-liquid solution in the vicinity of the Mott-transition.  
At $E_z=0$, one of the key bandstructure inputs is the degeneracy tied to MX/XM sites, which is split by $\sim\delta$ from the energy of the MM site.  At $\nu=1$, and for the typical values of $t_{TH},~t_{HH},~U_H,~U_T,~V$, we find the bosons delocalized across all three orbitals in a superfluid phase at small values of $E_z$, thereby quenching any tendency towards fractionalization. The evolution of $\langle b\rangle$ summed over all three sites is shown in Fig.~\ref{fig3_boson}a, along with the occupations $n_b$ on each of the three sublattice sites in Fig.~\ref{fig3_boson}b. The chiral exchange term mediates attraction between the spinons, leading to an inter-valley and fully gapped extended $s-$wave singlet paired state. Note that the expectation value for the electron pairing operators (suppressing the spatial and orbital labels for simplicity) is given by,  $\langle c c \rangle=\langle b^2 \rangle \langle ff\rangle$, within the mean field approximation. When the bosons form a superfluid ($\langle b \rangle \ne 0$) and the spinons are paired ($\langle f f\rangle\neq0$), the electron pair correlation function has a finite expectation value, and the resulting state is an electronic superconductor. On the other hand, across a Mott transition where $\langle b\rangle=0$, if the spinon pairing is not lost, the resulting state of gapped charge and paired spinons yields a fully gapped quantum spin liquid \cite{Senthil_SC_spinon,Grover_SC_spinon}. With increasing displacement field, the energies of the MX/XM sites are no longer degenerate, and split by the field, and we find that the holon and spinon localize only on the MX/MM sites, as shown in Fig.~\ref{fig:occupation}b. Note that the holon occupations are shared between the MX/MM sites as $1-\epsilon,~\epsilon$, respectively, with $\epsilon>0$; see e.g. green $\times$-markers in Fig.~\ref{fig3_boson} a and b. We find that the $\pm\epsilon-$fractions of the holons experience a net attraction due to the Coulomb interaction, $V$, which leads to a holon binding and an excitonic Mott insulator \cite{DCNC,SKDC24}. Note that the exciton here is formed between the positive and negatively charged holons, with compensated densities $\pm\epsilon$. This phase is stabilized in the strong $U$ and $V$ limit.
Importantly, the pattern of charge localization still implies an underlying layer-hybridized state. The spinon pairing also survives leading to an insulator with both spin and charge-gap; see green $\times$-marker in Fig.~\ref{fig4_spinon}a, b. This is the promised fully-gapped quantum spin liquid insulator. Further increasing the displacement-field, we find that the spinon occupations on the different sublattice sites change rapidly, with $\langle f^\dagger f\rangle_{\tn{MM}}\rightarrow0$; see orange {$\times$-marker in Figs.~\ref{fig4_spinon}a, b. The nature of the chiral exchange interaction (between the MM and MX/XM sites) is such that this automatically also leads to a loss of spinon pairing, without affecting the charge-localization in the Mott insulator. Thus, within this scenario, at large fields, there is a quantum phase transition between the LH spin-gapped Mott insulator to a layer polarized Mott insulator with the charge and spinons localized on the triangular lattice sites consisting of the MX-sites, as discussed in the Supplementary Material Sec.~II. Within our mean-field ansatz, this can naturally yield a spin-liquid with spinon Fermi surface, but could also yield more conventional ordered states with a finite expectation value for $\langle f^\dagger \vec\sigma f\rangle\neq0$ at specific ordering wavevectors. We leave a detailed study of competition between such states for future work. An unbiased determination of the energetics for the present model can only arise from numerically exact methods, which are beyond the scope of present work.

\begin{figure}[pth!]
\centering
\includegraphics[width=1.0\linewidth]{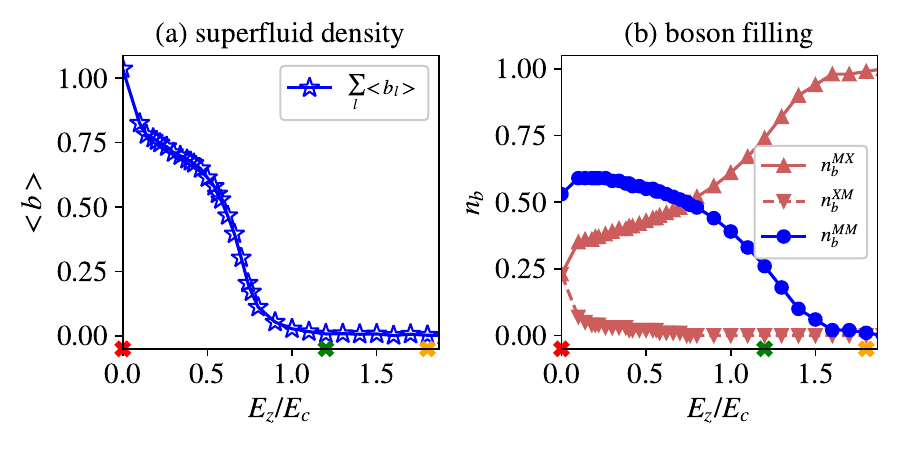} 
\caption{{\bf Evolution of the holon properties across the superfluid-Mott insulator transition}. (a) The holon condensate $\langle b\rangle$, which is related to the holon superfluid density, summed over the three orbitals, and (b) boson occupation as a function of displacement field. Red and green $\times$-markers correspond to the superconductor and gapped spin liquid illustrated in Fig.~\ref{fig:occupation}(a) and (b), respectively. Orange $\times$-marker represents a layer-polarized Mott insulator with the spinons and holons localized on the MX sites, at much higher fields; see the supplementary material Sec.~\ref{sec:SI_LPMI}. The model parameters used for all of our simulations are: $\delta=-14.9\tn{meV}$, $t_{TH}^{\left(1\right)}=10.78\tn{meV}$, $t_{HH}^{\left(1\right)}=0.55\tn{meV}$ , $t_{TT}^{\left(1\right)}=-1.95\tn{meV}$, $t_{TH}^{\left(2\right)}=-1.21\tn{meV}$, $t_{HH}^{\left(3\right)}=5.4\tn{meV}$, $U_H=35\tn{meV}$, $U_T=20\tn{meV}$, $V=40\tn{meV}$, and $J=10\tn{meV}$, respectively.}  
\label{fig3_boson}
\end{figure}

\begin{figure}[pth!]
\centering
\includegraphics[width=1.0\linewidth]{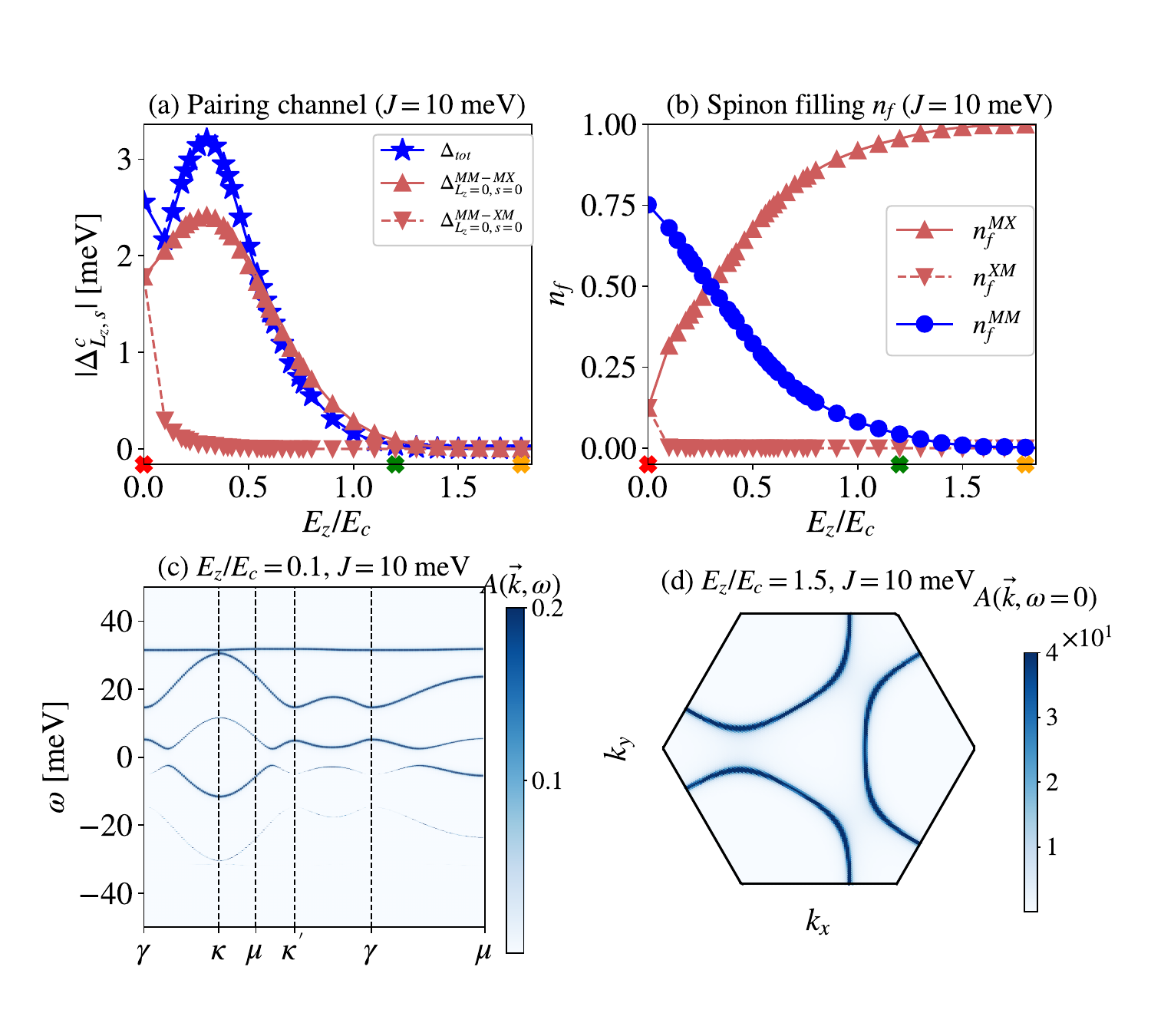} 
\caption{{\bf Evolution of the spinon properties with increasing displacement-field}. (a) Spinon pairing-gap $\Delta_{\tn{tot}}$, obtained from parton mean-field computation as a function of displacement-field, $E_z$. The bosonic Mott critical field is denoted by $E_c$. Slightly beyond $E_c$, the spinon sector remains paired. (b) Evolution of spinon occupation with increasing displacement-field. For $E_z \gtrsim E_c$ (green $\times$-marker), the spinon remains paired and layer-hybridized; for $E_z\gg E_c$ (orange $\times$-marker), the spinon sector tends to layer polarize. (c) Spinon Bogoliubov spectral function $A_f(\vec{k},\omega)$, evaluated at $E_z/E_c = 0.1$ and $J=10\tn{meV}$. A spinon gap $\sim 2 \tn{meV}$ is observed for the half-filled lowest band. (d) Spinon Bogoliubov spectral function $A_f(\vec{k},\omega = 0)$ at $E_z/E_c = 1.4$ and $J=10\tn{meV}$ within moir\'e Brillouin zone. The spinon gap $|\Delta_{\tn{tot}}|$ has been fully suppressed and $A(\vec{k},\omega = 0)$ shows the spinon Fermi surface. Model parameters are the same as in Fig.\ref{fig3_boson}} 
\label{fig4_spinon}
\end{figure}

In Fig.~\ref{fig3_boson}, we plot the evolution of the condensation expectation value, $\langle b\rangle$, summed over all the orbitals, which sets the superfluid density for  the holons, and holon-densities with increasing displacement-field, obtained from the parton mean-field computation using $H_b$. As noted above, at small displacement fields, the holons are clearly in a superfluid state (corresponding to Fig.~\ref{fig:occupation}a), with $\langle b\rangle_\ell\neq0~\forall\ell$. On the other hand, beyond a critical $E_z$, numerically we find $\langle b^\dagger b\rangle \neq 0$ for nearest-neighbor MX/MM sites and $\langle b\rangle\rightarrow0$. 
This is the Mott insulating phase (corresponding to Fig.~\ref{fig:occupation}b), which remains layer-hybridized. Eventually, with increasing $E_z$, the bosons favor layer-polarization (see Supplementary Material Sec.~II), where the system effectively becomes a triangular lattice system constructed out of the MX-sites.  Simultaneously, it is useful to track the evolution of the spinon-densities with increasing displacement-fields; see Fig.~\ref{fig:occupation} (bottom-row). At small displacement fields, the spinon occupations $\langle f^\dagger f\rangle_\ell\neq0~\forall\ell$ (Fig.~\ref{fig:occupation}a) in a spinon-metallic-like regime, which is unstable to pairing due to the chiral-exchange interaction; see Fig.~\ref{fig4_spinon}a. For $J>0$, attraction is mediated in the inter-valley, spin-singlet channel as noted above. Note that we are technically not including the contributions from the gauge-field fluctuations beyond mean-field theory here, which can suppress the pairing tendency as a result of standard amperean effects \cite{metlitski15}; we proceed with the assumption that the tendency towards spinon pairing remains prevalent. 

Within the same self-consistent parton mean field theory computations (Supplementary Material Sec.~I); the evolution of the angular-momentum ($L_z$)-resolved spinon pairing-gap with increasing displacement-field is shown in Fig.~\ref{fig4_spinon}(a). The angular momentum $L_z$ is defined as the phase winding of the spinon Cooper pairs which live on the three bonds connecting the $T-H$ sites. For arbitrary $E_z$, only $L_z = 0$ (extended s-wave) and $s=0$ (spin-singlet) channels develop a finite expectation value. At $E_z=0$, pairing between MM-MX and MM-XM are on an equal footing due to the presence of inversion symmetry. When $0<E_z<E_c$, a layer-imbalance develops but the spinon pairing gap $|\Delta_{\tn{tot}}|$ remains finite; for $E_z\gtrsim E_c$, the spinon pairing gap still remains finite whereas the boson is in Mott insulating phase (Fig. \ref{fig3_boson}a), namely as Z$_2$ spin liquid. When further increasing $E_z$, the pairing is fully suppressed by $E_z$, and the spinon sector tends to form a layer-polarized (LP) spinon-metal (Fig.~\ref{fig4_spinon}b), which can become unstable to other forms of orders. These features are reflected in the spinon spectral function, $A_f(\k,\omega)$, in the LH paired (Fig.~\ref{fig4_spinon}c) and LP unpaired (Fig.~\ref{fig4_spinon}d) regimes, respectively. Note that we have intentionally refrained from quoting the absolute values of $E_z$ in our analysis, as the values of the layer-polarization susceptibility for both of the matter fields is a priori unknown. 

Clearly, the nature of the resulting many-body phase is determined by the combination of the bosonic and fermionic correlators, respectively. As was noted above, when $\langle b\rangle\neq0$ in the superfluid phase, the resulting state is a superconductor as long as spinon-pairing is present (Fig.~\ref{fig:occupation}a). As a function of increasing displacement-field, there can be a tendency towards spontaneously broken $C_3-$symmetry \cite{constantin}, as well as other broken symmetries that include long-range magnetic order. If the holons remain condensed, and in the absence of the magnetic order, the resulting state is a nematic superconductor, whereas if the critical field for $C_3-$breaking is larger than the Mott transition for the holons, the nematicity onsets only in the gapped spin-liquid insulator.  Within our present scenario, when $\langle b\rangle=0$ in the Mott-insulating phase, for small displacement-fields the ground-state is an electrical insulator with both a charge and spin-gap, respectively. When the spinon pairing is lost, the system transitions into a Mott insulator with a charge-gap but no spin-gap, reminiscent of previous experiments in AA-stacked heterobilayers \cite{li2021continuous}. 

\section{Discussion}
 The intriguing phenomenology tied to the recently discovered continuous superconductor-to-insulator transition at a fixed commensurate filling in tWSe$_2$ has naturally led us to a scenario where the origin of fermionic pairing (due to spinons) lies in the insulator, and the electrical response (due to holons) is determined purely by the interplay of Hubbard interactions and charge-transfer gap between the different orbitals. Our proposal already motivates the need for a number of future experiments, which will be crucial for developing deeper theoretical insights into this problem. 

First and foremost, the temperature dependence of the magnetic susceptibility in the insulating phase using MCD (at low temperature) will help reveal whether a spin-gap exists along with the charge-gap. This has revealed unparalleled insights in a previous experiment in a Mott-insulator \cite{li2021continuous}. The above scenario suggests that in the insulator across the transition from the superconductor, the asymptotic low-temperature susceptibility will be exponentially suppressed. It is also plausible, based in part on our computations, that there are two distinct insulators, separated by a spin-gap closing transition (without any closing of the charge-gap) that can be distinguished at the lowest temperatures based on the MCD data. The kink-like feature in the insulating region in the experiment \cite{TMDSC} might be indicative of such a transition. 

It is worth addressing the possibility of other competing insulating states. For strongly interacting Hubbard-type models with geometrical frustrations, conventional antiferromagnetic orders are often more favored \cite{ssrev}. However, our focus here has been on the physics near the Mott-transition, where the local-moment physics is not entirely developed. Such weak Mott insulators with a small charge-gap are ideally suited for realizing candidate quantum spin liquid states \cite{senthil_08}. The state obtained here just across the transition, is an ideal parent state for the presumably unconventional superconductor. We note that a recent DMRG study for a much simpler, but classic, triangular lattice Hubbard model at fixed filling also found an exotic gapped spin liquid (in the bulk) across a metal-insulator transition \cite{Szasz_CSL_PRX}; the transition to the conventional magnetically ordered state only occurs deeper inside the insulating state. Establishing the full energy-landscape for the present multi-orbital (topological) model in the intermediate to strong-coupling regime is clearly beyond the realm of a (parton) mean-field theory setup and remains an exciting frontier for both exact and approximate numerical methods. We also note in passing that given the continuous nature of the transition, and in the absence of any fine tuning, any broken translation symmetries in the Mott-insulator proximate to the transition will be present in the superconductor; an alternative and more exotic scenario is that the critical point is deconfined, as has also been suggested in recent work \cite{Cenke_DQCP}. 

While our work serves as a useful starting point to study the relation between the experimental puzzles and our current understanding of the actual bandstructures in twisted homobilayers, the complexity associated with the three-orbital model and its associated topological character reveal a number of unique challenges. In future work and for complementary insights into the problem, it would be interesting to step away from the complicated microscopic details of the present setup, and describe a simpler version of the model with the interactions projected to the relevant low-energy topological bands directly in momentum-space. This can enable a better and universal understanding of the phase transition between a paramagnetic Mott insulator, including the fully-gapped spin liquid considered here, and a superconductor in a model of $C=\pm1$ Chern bands in a time-reversal symmetric setting at integer filling. A variational approach with the analog of a Gutzwiller-type wavefunction \cite{Grover_SC_spinon}, but implemented in momentum-space, would be an interesting first step.

We end by noting that one of the most exciting open questions is related to the nature of the metal-to-metal transition that occurs as a function of filling across the $\nu=1$ orders at a fixed displacement-field. Approaching from $\nu\rightarrow1^+$, a renormalized Fermi liquid with an increasing effective mass transitions either into an insulator, or a superconductor in the near vicinity of $\nu=1$. For $\nu\rightarrow1^-$, the properties of the metallic state are not entirely clear at present, but there are marked phenomenological differences from a conventional Fermi liquid. Further studies of these metallic phases, incorporating also the effects of disorder, might lead to an improved understanding of the global low-temperature phenomenology in the vicinity of $\nu=1$ in tWSe$_2$. In the present setup, we find that superconductivity persists over a small range of fillings near $\nu=1$; see Supplementary Material Sec.~III. Within the current scenario, it is worth noting that the superconducting $T_c$ is controlled by the phase-stiffness and not the pairing gap, which can be small both at $\nu=1$ (e.g. due to disorder effects \cite{SFMI} and suppressed tendency towards pairing \cite{metlitski15,TSPAL}) and with doping. Moreover, given the spinonic origin of the pairing, doping away potentially leads to a dramatic suppression of this tendency, when the spinons are prone to confinement. Investigating these effects in more careful detail and utilizing more sophisticated methods remains an interesting open problem.

{\it Note added.-} A related manuscript has also recently analyzed the pairing instabilities of spinons due to the chiral-exchange interaction \cite{Valentin_spinon_pairing}.
\section{Methods}
\label{sec:Methods}

{\bf Three-orbital model}
The non-interacting three-orbital model in the main text is given by 
\begin{subequations}
\beq   H_{\uparrow} &=& \sum_{\mathbf{k},\tau}c_{\mathbf{k},\tau}^{\dagger}\left[h_{1,\tau}\left(\mathbf{k}\right)+h_{2,\tau}\left(\mathbf{k}\right)\right]c_{\mathbf{k},\tau},~\tn{where} \label{ho}\\
h_{1}\left(\mathbf{k}\right) &=& \left(\begin{array}{ccc}
E_{z}-\mu & t_{TH}^{\left(1\right)}g_{\mathbf{k}} & t_{HH}^{\left(1\right)}f_{\mathbf{k}}^{*}\\
t_{TH}^{\left(1\right)}g_{\mathbf{k}}^{*} & -\delta-\mu & -t_{TH}^{\left(1\right)}g_{-\mathbf{k}}^{*}\\
t_{HH}^{\left(1\right)}f_{\mathbf{k}} & -t_{TH}^{\left(1\right)}g_{-\mathbf{k}} & -E_{z}-\mu
\end{array}\right),\\
h_{2}\left(\mathbf{k}\right) &=& \left(\begin{array}{ccc}
0 & -t_{TH}^{\left(2\right)}g_{-2\mathbf{k}} & t_{HH}^{\left(3\right)}f_{2\mathbf{k}}\\
-t_{TH}^{\left(2\right)}g_{-2\mathbf{k}}^{*} & t_{TT}^{\left(1\right)}h_{\mathbf{k}} & t_{TH}^{\left(2\right)}g_{2\mathbf{k}}^{*}\\
 t_{HH}^{\left(3\right)}f^{*}_{2\mathbf{k}} & t_{TH}^{\left(2\right)}g_{2\mathbf{k}} & 0
\end{array}\right).
\eeq    
\end{subequations}
The displacement-field, $E_z$, modifies the on-site energies and $\mu$ is the chemical potential. The hopping matrix-elements, $t_{ab}$ ($a, b\equiv T, H$), for the triangular and honeycomb lattice sites are shown in Fig.~\ref{fig:model}(a). The associated momentum-space form-factors are defined as, 
\begin{subequations}
\beq
f_{\mathbf{k}}	&=&\sum_{j=0,1,2}e^{i\mathbf{k}\cdot\mathbf{u}_{j}},~~g_{\mathbf{k}}=\sum_{j=0,1,2}e^{i2\pi\left(j-1\right)/3}e^{i\mathbf{k}\cdot\mathbf{u}_{j}}, \\
h_{\mathbf{k}}&=&2\sum_{j=1,2,3}\cos\left(\mathbf{k}\cdot\mathbf{a}_{j}\right),
\eeq   
\end{subequations}
where $\vec{a}_1=(a_M,0)$, $\vec{a}_j=C_3^{j-1}\vec{a}_1$, $\vec{u}_0=(\vec{a}_1-\vec{a}_2)/3$, $\vec{u}_j=C_3^j\vec{u}_0$, and $a_M$ is the moire lattice constant. The specific orbital content is that of orbitals localized at the XM/MX ($H$) and MM ($T$) stacking regions of the bilayer with an s-wave character. The former ($H$) are layer polarized as interlayer tuneling vanishes at these stacking regions, while the latter ($T$) is layer hybridized since MM stacking regions map to themselves under $C_{2y}$ symmetry, implying that they possess mixed character from both layers \cite{valentin24}.

{\bf Parton mean-field theory} The parton representation proceeds in the usual fashion \cite{Florens_PRB_2004,Zhao_PRB_2007}, where we express $c_{\r,\l,\sigma} = b_{\r,\l} f_{\r,\l,\sigma}$, with the $b_{\r,\l}$ representing spinless charged holon fields at site $\r$ with orbital $\l$, and the $f_{\r,\l,\sigma}$ denoting spinful neutral spinons that also carry the orbital index $\l$. The holon annihilation operator can be written in terms of a quantum rotor representation, $b_{\vec{r},\ell}=e^{i\theta_{\vec{r},\ell}}$, which raises the rotor charge $n^{\theta}_{\vec{r},\ell}$ by 1. The local constraint that helps project the problem back to the physical Hilbert space is given by  $\langle \sum_\l n_{\r,\l}^\theta \rangle + \langle \sum_{\l,\sigma} n^f_{\r,\l,\sigma}\rangle = \tn{const}$. Note that the microscopic Hamiltonian does not forbid double occupancy, and we will implement the filling constraints for the parton fields on average in our numerical simulations.

The mean-field Hamiltonian takes the form, $H_{\rm{MF}} =   H_b(\{\chi\}) + H_f(\{\chi,\Delta,B\})$, where the variational parameters in the matter field sectors are tied to the correlators, $B_{\r\r'}^{\l\l'}\equiv \langle b^\dagger_{\r,\l} b^{\phantom\dagger}_{\r',\l'} \rangle$, $\chi_{\r\r',\sigma}^{\l\l'}\equiv \langle f_{\r,\l,\sigma}^\dagger f^{\phantom\dagger}_{\r',\l',\sigma}\rangle$, and $\Delta_{\r\r',\sigma\sigma'}^{\l\l'}\equiv \langle \ve_{\sigma\sigma'}f_{\r,\l,\sigma}f_{\r',\l',\sigma'}\rangle$, respectively. Clearly, the $B-$correlator is evaluated with respect to $H_b$ and renormalizes the spinon bandwidth in $H_f$. Similarly, $\chi$ is evaluated with respect to $H_f$, which arises both from the bare bandwidth and the chiral-exchange term, and renormalizes the boson hoppings in $H_b$, as well as the spinon hoppings in $H_f$. Finally, $\Delta$ is also evaluated with respect to $H_f$ and drives the pairing of spinons. To deal with $H_b$, we utilize a 3-site cluster approximation comprising all the three orbitals, within which we impose the global $U(1)$ number conservation. Given that $t_{TH}^{(1)}$ is the dominant hopping, for the computations presented here, we consider the equilateral triangles comprising nearest-neighbor orbitals within the clusters; see Supplementary Material Sec.~IV for additional details and a more elaborate choice of clusters. 

\textit{Data availability.-} The data generated in this study and supporting the manuscript figures including those in the Supplementary Information have been deposited in the Zenodo database https://doi.org/10.5281/zenodo.14733332

\textit{Code availability.-} The standard codes for the dataset generated in the current study are available from the corresponding author upon request.

\bibliographystyle{apsrev4-1_custom}
\bibliography{refs}

\begin{thebibliography}{63}%
\makeatletter
\providecommand \@ifxundefined [1]{%
 \@ifx{#1\undefined}
}%
\providecommand \@ifnum [1]{%
 \ifnum #1\expandafter \@firstoftwo
 \else \expandafter \@secondoftwo
 \fi
}%
\providecommand \@ifx [1]{%
 \ifx #1\expandafter \@firstoftwo
 \else \expandafter \@secondoftwo
 \fi
}%
\providecommand \natexlab [1]{#1}%
\providecommand \enquote  [1]{``#1''}%
\providecommand \bibnamefont  [1]{#1}%
\providecommand \bibfnamefont [1]{#1}%
\providecommand \citenamefont [1]{#1}%
\providecommand \href@noop [0]{\@secondoftwo}%
\providecommand \href [0]{\begingroup \@sanitize@url \@href}%
\providecommand \@href[1]{\@@startlink{#1}\@@href}%
\providecommand \@@href[1]{\endgroup#1\@@endlink}%
\providecommand \@sanitize@url [0]{\catcode `\\12\catcode `\$12\catcode `\&12\catcode `\#12\catcode `\^12\catcode `\_12\catcode `\%12\relax}%
\providecommand \@@startlink[1]{}%
\providecommand \@@endlink[0]{}%
\providecommand \url  [0]{\begingroup\@sanitize@url \@url }%
\providecommand \@url [1]{\endgroup\@href {#1}{\urlprefix }}%
\providecommand \urlprefix  [0]{URL }%
\providecommand \Eprint [0]{\href }%
\providecommand \doibase [0]{http://dx.doi.org/}%
\providecommand \selectlanguage [0]{\@gobble}%
\providecommand \bibinfo  [0]{\@secondoftwo}%
\providecommand \bibfield  [0]{\@secondoftwo}%
\providecommand \translation [1]{[#1]}%
\providecommand \BibitemOpen [0]{}%
\providecommand \bibitemStop [0]{}%
\providecommand \bibitemNoStop [0]{.\EOS\space}%
\providecommand \EOS [0]{\spacefactor3000\relax}%
\providecommand \BibitemShut  [1]{\csname bibitem#1\endcsname}%
\let\auto@bib@innerbib\@empty
\bibitem [{\citenamefont {Cao}\ \emph {et~al.}(2018{\natexlab{a}})\citenamefont {Cao}, \citenamefont {Fatemi}, \citenamefont {Fang}, \citenamefont {Watanabe}, \citenamefont {Taniguchi}, \citenamefont {Kaxiras},\ and\ \citenamefont {Jarillo-Herrero}}]{Cao2018b}%
  \BibitemOpen
  \bibfield  {author} {\bibinfo {author} {\bibfnamefont {Y.}~\bibnamefont {Cao}}, \bibinfo {author} {\bibfnamefont {V.}~\bibnamefont {Fatemi}}, \bibinfo {author} {\bibfnamefont {S.}~\bibnamefont {Fang}}, \bibinfo {author} {\bibfnamefont {K.}~\bibnamefont {Watanabe}}, \bibinfo {author} {\bibfnamefont {T.}~\bibnamefont {Taniguchi}}, \bibinfo {author} {\bibfnamefont {E.}~\bibnamefont {Kaxiras}}, \ and\ \bibinfo {author} {\bibfnamefont {P.}~\bibnamefont {Jarillo-Herrero}},\ }\bibfield  {title} {\enquote {\bibinfo {title} {Unconventional superconductivity in magic-angle graphene superlattices},}\ }\href {\doibase 10.1038/nature26160} {\bibfield  {journal} {\bibinfo  {journal} {Nature}\ }\textbf {\bibinfo {volume} {556}},\ \bibinfo {pages} {43} (\bibinfo {year} {2018}{\natexlab{a}})}\BibitemShut {NoStop}%
\bibitem [{\citenamefont {Yankowitz}\ \emph {et~al.}(2019)\citenamefont {Yankowitz}, \citenamefont {Chen}, \citenamefont {Polshyn}, \citenamefont {Zhang}, \citenamefont {Watanabe}, \citenamefont {Taniguchi}, \citenamefont {Graf}, \citenamefont {Young},\ and\ \citenamefont {Dean}}]{Yankowitz_2019}%
  \BibitemOpen
  \bibfield  {author} {\bibinfo {author} {\bibfnamefont {M.}~\bibnamefont {Yankowitz}}, \bibinfo {author} {\bibfnamefont {S.}~\bibnamefont {Chen}}, \bibinfo {author} {\bibfnamefont {H.}~\bibnamefont {Polshyn}}, \bibinfo {author} {\bibfnamefont {Y.}~\bibnamefont {Zhang}}, \bibinfo {author} {\bibfnamefont {K.}~\bibnamefont {Watanabe}}, \bibinfo {author} {\bibfnamefont {T.}~\bibnamefont {Taniguchi}}, \bibinfo {author} {\bibfnamefont {D.}~\bibnamefont {Graf}}, \bibinfo {author} {\bibfnamefont {A.~F.}\ \bibnamefont {Young}}, \ and\ \bibinfo {author} {\bibfnamefont {C.~R.}\ \bibnamefont {Dean}},\ }\bibfield  {title} {\enquote {\bibinfo {title} {Tuning superconductivity in twisted bilayer graphene},}\ }\href {\doibase 10.1126/science.aav1910} {\bibfield  {journal} {\bibinfo  {journal} {Science}\ }\textbf {\bibinfo {volume} {363}},\ \bibinfo {pages} {1059} (\bibinfo {year} {2019})}\BibitemShut {NoStop}%
\bibitem [{\citenamefont {Lu}\ \emph {et~al.}(2019)\citenamefont {Lu}, \citenamefont {Stepanov}, \citenamefont {Yang}, \citenamefont {Xie}, \citenamefont {Aamir}, \citenamefont {Das}, \citenamefont {Urgell}, \citenamefont {Watanabe}, \citenamefont {Taniguchi}, \citenamefont {Zhang}, \citenamefont {Bachtold}, \citenamefont {MacDonald},\ and\ \citenamefont {Efetov}}]{Lu2019}%
  \BibitemOpen
  \bibfield  {author} {\bibinfo {author} {\bibfnamefont {X.}~\bibnamefont {Lu}}, \bibinfo {author} {\bibfnamefont {P.}~\bibnamefont {Stepanov}}, \bibinfo {author} {\bibfnamefont {W.}~\bibnamefont {Yang}}, \bibinfo {author} {\bibfnamefont {M.}~\bibnamefont {Xie}}, \bibinfo {author} {\bibfnamefont {M.~A.}\ \bibnamefont {Aamir}}, \bibinfo {author} {\bibfnamefont {I.}~\bibnamefont {Das}}, \bibinfo {author} {\bibfnamefont {C.}~\bibnamefont {Urgell}}, \bibinfo {author} {\bibfnamefont {K.}~\bibnamefont {Watanabe}}, \bibinfo {author} {\bibfnamefont {T.}~\bibnamefont {Taniguchi}}, \bibinfo {author} {\bibfnamefont {G.}~\bibnamefont {Zhang}}, \bibinfo {author} {\bibfnamefont {A.}~\bibnamefont {Bachtold}}, \bibinfo {author} {\bibfnamefont {A.~H.}\ \bibnamefont {MacDonald}}, \ and\ \bibinfo {author} {\bibfnamefont {D.~K.}\ \bibnamefont {Efetov}},\ }\bibfield  {title} {\enquote {\bibinfo {title} {Superconductors, orbital magnets and correlated states in magic-angle bilayer graphene},}\ }\href {\doibase
  10.1038/s41586-019-1695-0} {\bibfield  {journal} {\bibinfo  {journal} {Nature}\ }\textbf {\bibinfo {volume} {574}},\ \bibinfo {pages} {653} (\bibinfo {year} {2019})}\BibitemShut {NoStop}%
\bibitem [{\citenamefont {Arora}\ \emph {et~al.}(2020)\citenamefont {Arora}, \citenamefont {Polski}, \citenamefont {Zhang}, \citenamefont {Thomson}, \citenamefont {Choi}, \citenamefont {Kim}, \citenamefont {Lin}, \citenamefont {Wilson}, \citenamefont {Xu}, \citenamefont {Chu}, \citenamefont {Watanabe}, \citenamefont {Taniguchi}, \citenamefont {Alicea},\ and\ \citenamefont {Nadj-Perge}}]{Arora2020SC}%
  \BibitemOpen
  \bibfield  {author} {\bibinfo {author} {\bibfnamefont {H.~S.}\ \bibnamefont {Arora}}, \bibinfo {author} {\bibfnamefont {R.}~\bibnamefont {Polski}}, \bibinfo {author} {\bibfnamefont {Y.}~\bibnamefont {Zhang}}, \bibinfo {author} {\bibfnamefont {A.}~\bibnamefont {Thomson}}, \bibinfo {author} {\bibfnamefont {Y.}~\bibnamefont {Choi}}, \bibinfo {author} {\bibfnamefont {H.}~\bibnamefont {Kim}}, \bibinfo {author} {\bibfnamefont {Z.}~\bibnamefont {Lin}}, \bibinfo {author} {\bibfnamefont {I.~Z.}\ \bibnamefont {Wilson}}, \bibinfo {author} {\bibfnamefont {X.}~\bibnamefont {Xu}}, \bibinfo {author} {\bibfnamefont {J.-H.}\ \bibnamefont {Chu}}, \bibinfo {author} {\bibfnamefont {K.}~\bibnamefont {Watanabe}}, \bibinfo {author} {\bibfnamefont {T.}~\bibnamefont {Taniguchi}}, \bibinfo {author} {\bibfnamefont {J.}~\bibnamefont {Alicea}}, \ and\ \bibinfo {author} {\bibfnamefont {S.}~\bibnamefont {Nadj-Perge}},\ }\bibfield  {title} {\enquote {\bibinfo {title} {Superconductivity in metallic twisted bilayer graphene stabilized by
  {WSe}2},}\ }\href {\doibase 10.1038/s41586-020-2473-8} {\bibfield  {journal} {\bibinfo  {journal} {Nature}\ }\textbf {\bibinfo {volume} {583}},\ \bibinfo {pages} {379} (\bibinfo {year} {2020})}\BibitemShut {NoStop}%
\bibitem [{\citenamefont {Hao}\ \emph {et~al.}(2021)\citenamefont {Hao}, \citenamefont {Zimmerman}, \citenamefont {Ledwith}, \citenamefont {Khalaf}, \citenamefont {Najafabadi}, \citenamefont {Watanabe}, \citenamefont {Taniguchi}, \citenamefont {Vishwanath},\ and\ \citenamefont {Kim}}]{Hao2021TTGelectric}%
  \BibitemOpen
  \bibfield  {author} {\bibinfo {author} {\bibfnamefont {Z.}~\bibnamefont {Hao}}, \bibinfo {author} {\bibfnamefont {A.~M.}\ \bibnamefont {Zimmerman}}, \bibinfo {author} {\bibfnamefont {P.}~\bibnamefont {Ledwith}}, \bibinfo {author} {\bibfnamefont {E.}~\bibnamefont {Khalaf}}, \bibinfo {author} {\bibfnamefont {D.~H.}\ \bibnamefont {Najafabadi}}, \bibinfo {author} {\bibfnamefont {K.}~\bibnamefont {Watanabe}}, \bibinfo {author} {\bibfnamefont {T.}~\bibnamefont {Taniguchi}}, \bibinfo {author} {\bibfnamefont {A.}~\bibnamefont {Vishwanath}}, \ and\ \bibinfo {author} {\bibfnamefont {P.}~\bibnamefont {Kim}},\ }\bibfield  {title} {\enquote {\bibinfo {title} {Electric field{\textendash}tunable superconductivity in alternating-twist magic-angle trilayer graphene},}\ }\href {\doibase 10.1126/science.abg0399} {\bibfield  {journal} {\bibinfo  {journal} {Science}\ }\textbf {\bibinfo {volume} {371}},\ \bibinfo {pages} {1133} (\bibinfo {year} {2021})}\BibitemShut {NoStop}%
\bibitem [{\citenamefont {Oh}\ \emph {et~al.}(2021)\citenamefont {Oh}, \citenamefont {Nuckolls}, \citenamefont {Wong}, \citenamefont {Lee}, \citenamefont {Liu}, \citenamefont {Watanabe}, \citenamefont {Taniguchi},\ and\ \citenamefont {Yazdani}}]{Oh2021unconventional}%
  \BibitemOpen
  \bibfield  {author} {\bibinfo {author} {\bibfnamefont {M.}~\bibnamefont {Oh}}, \bibinfo {author} {\bibfnamefont {K.~P.}\ \bibnamefont {Nuckolls}}, \bibinfo {author} {\bibfnamefont {D.}~\bibnamefont {Wong}}, \bibinfo {author} {\bibfnamefont {R.~L.}\ \bibnamefont {Lee}}, \bibinfo {author} {\bibfnamefont {X.}~\bibnamefont {Liu}}, \bibinfo {author} {\bibfnamefont {K.}~\bibnamefont {Watanabe}}, \bibinfo {author} {\bibfnamefont {T.}~\bibnamefont {Taniguchi}}, \ and\ \bibinfo {author} {\bibfnamefont {A.}~\bibnamefont {Yazdani}},\ }\bibfield  {title} {\enquote {\bibinfo {title} {Evidence for unconventional superconductivity in twisted bilayer graphene},}\ }\href {\doibase 10.1038/s41586-021-04121-x} {\bibfield  {journal} {\bibinfo  {journal} {Nature}\ }\textbf {\bibinfo {volume} {600}},\ \bibinfo {pages} {240} (\bibinfo {year} {2021})}\BibitemShut {NoStop}%
\bibitem [{\citenamefont {Zhou}\ \emph {et~al.}(2021{\natexlab{a}})\citenamefont {Zhou}, \citenamefont {Xie}, \citenamefont {Taniguchi}, \citenamefont {Watanabe},\ and\ \citenamefont {Young}}]{Zhou2021RTGSC}%
  \BibitemOpen
  \bibfield  {author} {\bibinfo {author} {\bibfnamefont {H.}~\bibnamefont {Zhou}}, \bibinfo {author} {\bibfnamefont {T.}~\bibnamefont {Xie}}, \bibinfo {author} {\bibfnamefont {T.}~\bibnamefont {Taniguchi}}, \bibinfo {author} {\bibfnamefont {K.}~\bibnamefont {Watanabe}}, \ and\ \bibinfo {author} {\bibfnamefont {A.~F.}\ \bibnamefont {Young}},\ }\bibfield  {title} {\enquote {\bibinfo {title} {Superconductivity in rhombohedral trilayer graphene},}\ }\href {\doibase 10.1038/s41586-021-03926-0} {\bibfield  {journal} {\bibinfo  {journal} {Nature}\ }\textbf {\bibinfo {volume} {598}},\ \bibinfo {pages} {434} (\bibinfo {year} {2021}{\natexlab{a}})}\BibitemShut {NoStop}%
\bibitem [{\citenamefont {Zhou}\ \emph {et~al.}(2022)\citenamefont {Zhou}, \citenamefont {Holleis}, \citenamefont {Saito}, \citenamefont {Cohen}, \citenamefont {Huynh}, \citenamefont {Patterson}, \citenamefont {Yang}, \citenamefont {Taniguchi}, \citenamefont {Watanabe},\ and\ \citenamefont {Young}}]{Zhou2022BBGSC}%
  \BibitemOpen
  \bibfield  {author} {\bibinfo {author} {\bibfnamefont {H.}~\bibnamefont {Zhou}}, \bibinfo {author} {\bibfnamefont {L.}~\bibnamefont {Holleis}}, \bibinfo {author} {\bibfnamefont {Y.}~\bibnamefont {Saito}}, \bibinfo {author} {\bibfnamefont {L.}~\bibnamefont {Cohen}}, \bibinfo {author} {\bibfnamefont {W.}~\bibnamefont {Huynh}}, \bibinfo {author} {\bibfnamefont {C.~L.}\ \bibnamefont {Patterson}}, \bibinfo {author} {\bibfnamefont {F.}~\bibnamefont {Yang}}, \bibinfo {author} {\bibfnamefont {T.}~\bibnamefont {Taniguchi}}, \bibinfo {author} {\bibfnamefont {K.}~\bibnamefont {Watanabe}}, \ and\ \bibinfo {author} {\bibfnamefont {A.~F.}\ \bibnamefont {Young}},\ }\bibfield  {title} {\enquote {\bibinfo {title} {Isospin magnetism and spin-polarized superconductivity in bernal bilayer graphene},}\ }\href {\doibase 10.1126/science.abm8386} {\bibfield  {journal} {\bibinfo  {journal} {Science}\ }\textbf {\bibinfo {volume} {375}},\ \bibinfo {pages} {774} (\bibinfo {year} {2022})}\BibitemShut {NoStop}%
\bibitem [{\citenamefont {Zhang}\ \emph {et~al.}(2023)\citenamefont {Zhang}, \citenamefont {Polski}, \citenamefont {Thomson}, \citenamefont {Lantagne-Hurtubise}, \citenamefont {Lewandowski}, \citenamefont {Zhou}, \citenamefont {Watanabe}, \citenamefont {Taniguchi}, \citenamefont {Alicea},\ and\ \citenamefont {Nadj-Perge}}]{Zhang2023BBGSC}%
  \BibitemOpen
  \bibfield  {author} {\bibinfo {author} {\bibfnamefont {Y.}~\bibnamefont {Zhang}}, \bibinfo {author} {\bibfnamefont {R.}~\bibnamefont {Polski}}, \bibinfo {author} {\bibfnamefont {A.}~\bibnamefont {Thomson}}, \bibinfo {author} {\bibfnamefont {{\'{E}}.}~\bibnamefont {Lantagne-Hurtubise}}, \bibinfo {author} {\bibfnamefont {C.}~\bibnamefont {Lewandowski}}, \bibinfo {author} {\bibfnamefont {H.}~\bibnamefont {Zhou}}, \bibinfo {author} {\bibfnamefont {K.}~\bibnamefont {Watanabe}}, \bibinfo {author} {\bibfnamefont {T.}~\bibnamefont {Taniguchi}}, \bibinfo {author} {\bibfnamefont {J.}~\bibnamefont {Alicea}}, \ and\ \bibinfo {author} {\bibfnamefont {S.}~\bibnamefont {Nadj-Perge}},\ }\bibfield  {title} {\enquote {\bibinfo {title} {Enhanced superconductivity in spin{\textendash}orbit proximitized bilayer graphene},}\ }\href {\doibase 10.1038/s41586-022-05446-x} {\bibfield  {journal} {\bibinfo  {journal} {Nature}\ }\textbf {\bibinfo {volume} {613}},\ \bibinfo {pages} {268} (\bibinfo {year} {2023})}\BibitemShut {NoStop}%
\bibitem [{\citenamefont {Cao}\ \emph {et~al.}(2018{\natexlab{b}})\citenamefont {Cao}, \citenamefont {Fatemi}, \citenamefont {Demir}, \citenamefont {Fang}, \citenamefont {Tomarken}, \citenamefont {Luo}, \citenamefont {Sanchez-Yamagishi}, \citenamefont {Watanabe}, \citenamefont {Taniguchi}, \citenamefont {Kaxiras}, \citenamefont {Ashoori},\ and\ \citenamefont {Jarillo-Herrero}}]{Cao2018}%
  \BibitemOpen
  \bibfield  {author} {\bibinfo {author} {\bibfnamefont {Y.}~\bibnamefont {Cao}}, \bibinfo {author} {\bibfnamefont {V.}~\bibnamefont {Fatemi}}, \bibinfo {author} {\bibfnamefont {A.}~\bibnamefont {Demir}}, \bibinfo {author} {\bibfnamefont {S.}~\bibnamefont {Fang}}, \bibinfo {author} {\bibfnamefont {S.~L.}\ \bibnamefont {Tomarken}}, \bibinfo {author} {\bibfnamefont {J.~Y.}\ \bibnamefont {Luo}}, \bibinfo {author} {\bibfnamefont {J.~D.}\ \bibnamefont {Sanchez-Yamagishi}}, \bibinfo {author} {\bibfnamefont {K.}~\bibnamefont {Watanabe}}, \bibinfo {author} {\bibfnamefont {T.}~\bibnamefont {Taniguchi}}, \bibinfo {author} {\bibfnamefont {E.}~\bibnamefont {Kaxiras}}, \bibinfo {author} {\bibfnamefont {R.~C.}\ \bibnamefont {Ashoori}}, \ and\ \bibinfo {author} {\bibfnamefont {P.}~\bibnamefont {Jarillo-Herrero}},\ }\bibfield  {title} {\enquote {\bibinfo {title} {Correlated insulator behaviour at half-filling in magic-angle graphene superlattices},}\ }\href {\doibase 10.1038/nature26154} {\bibfield  {journal} {\bibinfo
  {journal} {Nature}\ }\textbf {\bibinfo {volume} {556}},\ \bibinfo {pages} {80} (\bibinfo {year} {2018}{\natexlab{b}})}\BibitemShut {NoStop}%
\bibitem [{\citenamefont {Zhou}\ \emph {et~al.}(2021{\natexlab{b}})\citenamefont {Zhou}, \citenamefont {Xie}, \citenamefont {Ghazaryan}, \citenamefont {Holder}, \citenamefont {Ehrets}, \citenamefont {Spanton}, \citenamefont {Taniguchi}, \citenamefont {Watanabe}, \citenamefont {Berg}, \citenamefont {Serbyn},\ and\ \citenamefont {Young}}]{Zhou2021RTGhalf}%
  \BibitemOpen
  \bibfield  {author} {\bibinfo {author} {\bibfnamefont {H.}~\bibnamefont {Zhou}}, \bibinfo {author} {\bibfnamefont {T.}~\bibnamefont {Xie}}, \bibinfo {author} {\bibfnamefont {A.}~\bibnamefont {Ghazaryan}}, \bibinfo {author} {\bibfnamefont {T.}~\bibnamefont {Holder}}, \bibinfo {author} {\bibfnamefont {J.~R.}\ \bibnamefont {Ehrets}}, \bibinfo {author} {\bibfnamefont {E.~M.}\ \bibnamefont {Spanton}}, \bibinfo {author} {\bibfnamefont {T.}~\bibnamefont {Taniguchi}}, \bibinfo {author} {\bibfnamefont {K.}~\bibnamefont {Watanabe}}, \bibinfo {author} {\bibfnamefont {E.}~\bibnamefont {Berg}}, \bibinfo {author} {\bibfnamefont {M.}~\bibnamefont {Serbyn}}, \ and\ \bibinfo {author} {\bibfnamefont {A.~F.}\ \bibnamefont {Young}},\ }\bibfield  {title} {\enquote {\bibinfo {title} {Half- and quarter-metals in rhombohedral trilayer graphene},}\ }\href {\doibase 10.1038/s41586-021-03938-w} {\bibfield  {journal} {\bibinfo  {journal} {Nature}\ }\textbf {\bibinfo {volume} {598}},\ \bibinfo {pages} {429} (\bibinfo {year}
  {2021}{\natexlab{b}})}\BibitemShut {NoStop}%
\bibitem [{\citenamefont {Saito}\ \emph {et~al.}(2020)\citenamefont {Saito}, \citenamefont {Ge}, \citenamefont {Watanabe}, \citenamefont {Taniguchi},\ and\ \citenamefont {Young}}]{saito2020independent}%
  \BibitemOpen
  \bibfield  {author} {\bibinfo {author} {\bibfnamefont {Y.}~\bibnamefont {Saito}}, \bibinfo {author} {\bibfnamefont {J.}~\bibnamefont {Ge}}, \bibinfo {author} {\bibfnamefont {K.}~\bibnamefont {Watanabe}}, \bibinfo {author} {\bibfnamefont {T.}~\bibnamefont {Taniguchi}}, \ and\ \bibinfo {author} {\bibfnamefont {A.~F.}\ \bibnamefont {Young}},\ }\bibfield  {title} {\enquote {\bibinfo {title} {Independent superconductors and correlated insulators in twisted bilayer graphene},}\ }\href@noop {} {\bibfield  {journal} {\bibinfo  {journal} {Nature Physics}\ }\textbf {\bibinfo {volume} {16}},\ \bibinfo {pages} {926} (\bibinfo {year} {2020})}\BibitemShut {NoStop}%
\bibitem [{\citenamefont {Stepanov}\ \emph {et~al.}(2020)\citenamefont {Stepanov}, \citenamefont {Das}, \citenamefont {Lu}, \citenamefont {Fahimniya}, \citenamefont {Watanabe}, \citenamefont {Taniguchi}, \citenamefont {Koppens}, \citenamefont {Lischner}, \citenamefont {Levitov},\ and\ \citenamefont {Efetov}}]{Stepanov_2020}%
  \BibitemOpen
  \bibfield  {author} {\bibinfo {author} {\bibfnamefont {P.}~\bibnamefont {Stepanov}}, \bibinfo {author} {\bibfnamefont {I.}~\bibnamefont {Das}}, \bibinfo {author} {\bibfnamefont {X.}~\bibnamefont {Lu}}, \bibinfo {author} {\bibfnamefont {A.}~\bibnamefont {Fahimniya}}, \bibinfo {author} {\bibfnamefont {K.}~\bibnamefont {Watanabe}}, \bibinfo {author} {\bibfnamefont {T.}~\bibnamefont {Taniguchi}}, \bibinfo {author} {\bibfnamefont {F.~H.~L.}\ \bibnamefont {Koppens}}, \bibinfo {author} {\bibfnamefont {J.}~\bibnamefont {Lischner}}, \bibinfo {author} {\bibfnamefont {L.}~\bibnamefont {Levitov}}, \ and\ \bibinfo {author} {\bibfnamefont {D.~K.}\ \bibnamefont {Efetov}},\ }\bibfield  {title} {\enquote {\bibinfo {title} {Untying the insulating and superconducting orders in magic-angle graphene},}\ }\href {\doibase 10.1038/s41586-020-2459-6} {\bibfield  {journal} {\bibinfo  {journal} {Nature}\ }\textbf {\bibinfo {volume} {583}},\ \bibinfo {pages} {375} (\bibinfo {year} {2020})}\BibitemShut {NoStop}%
\bibitem [{\citenamefont {Liu}\ \emph {et~al.}(2021)\citenamefont {Liu}, \citenamefont {Wang}, \citenamefont {Watanabe}, \citenamefont {Taniguchi}, \citenamefont {Vafek},\ and\ \citenamefont {Li}}]{liu2021tuning}%
  \BibitemOpen
  \bibfield  {author} {\bibinfo {author} {\bibfnamefont {X.}~\bibnamefont {Liu}}, \bibinfo {author} {\bibfnamefont {Z.}~\bibnamefont {Wang}}, \bibinfo {author} {\bibfnamefont {K.}~\bibnamefont {Watanabe}}, \bibinfo {author} {\bibfnamefont {T.}~\bibnamefont {Taniguchi}}, \bibinfo {author} {\bibfnamefont {O.}~\bibnamefont {Vafek}}, \ and\ \bibinfo {author} {\bibfnamefont {J.}~\bibnamefont {Li}},\ }\bibfield  {title} {\enquote {\bibinfo {title} {Tuning electron correlation in magic-angle twisted bilayer graphene using coulomb screening},}\ }\href@noop {} {\bibfield  {journal} {\bibinfo  {journal} {Science}\ }\textbf {\bibinfo {volume} {371}},\ \bibinfo {pages} {1261} (\bibinfo {year} {2021})}\BibitemShut {NoStop}%
\bibitem [{\citenamefont {Xia}\ \emph {et~al.}(2024)\citenamefont {Xia}, \citenamefont {Han}, \citenamefont {Watanabe}, \citenamefont {Taniguchi}, \citenamefont {Shan},\ and\ \citenamefont {Mak}}]{TMDSC}%
  \BibitemOpen
  \bibfield  {author} {\bibinfo {author} {\bibfnamefont {Y.}~\bibnamefont {Xia}}, \bibinfo {author} {\bibfnamefont {Z.}~\bibnamefont {Han}}, \bibinfo {author} {\bibfnamefont {K.}~\bibnamefont {Watanabe}}, \bibinfo {author} {\bibfnamefont {T.}~\bibnamefont {Taniguchi}}, \bibinfo {author} {\bibfnamefont {J.}~\bibnamefont {Shan}}, \ and\ \bibinfo {author} {\bibfnamefont {K.~F.}\ \bibnamefont {Mak}},\ }\bibfield  {title} {\enquote {\bibinfo {title} {Superconductivity in twisted bilayer wse2},}\ }\href {\doibase 10.1038/s41586-024-08116-2} {\bibfield  {journal} {\bibinfo  {journal} {Nature}\ } (\bibinfo {year} {2024}),\ 10.1038/s41586-024-08116-2}\BibitemShut {NoStop}%
\bibitem [{\citenamefont {{Guo}}\ \emph {et~al.}(2024)\citenamefont {{Guo}}, \citenamefont {{Pack}}, \citenamefont {{Swann}}, \citenamefont {{Holtzman}}, \citenamefont {{Cothrine}}, \citenamefont {{Watanabe}}, \citenamefont {{Taniguchi}}, \citenamefont {{Mandrus}}, \citenamefont {{Barmak}}, \citenamefont {{Hone}}, \citenamefont {{Millis}}, \citenamefont {{Pasupathy}},\ and\ \citenamefont {{Dean}}}]{Dean24}%
  \BibitemOpen
  \bibfield  {author} {\bibinfo {author} {\bibfnamefont {Y.}~\bibnamefont {{Guo}}}, \bibinfo {author} {\bibfnamefont {J.}~\bibnamefont {{Pack}}}, \bibinfo {author} {\bibfnamefont {J.}~\bibnamefont {{Swann}}}, \bibinfo {author} {\bibfnamefont {L.}~\bibnamefont {{Holtzman}}}, \bibinfo {author} {\bibfnamefont {M.}~\bibnamefont {{Cothrine}}}, \bibinfo {author} {\bibfnamefont {K.}~\bibnamefont {{Watanabe}}}, \bibinfo {author} {\bibfnamefont {T.}~\bibnamefont {{Taniguchi}}}, \bibinfo {author} {\bibfnamefont {D.}~\bibnamefont {{Mandrus}}}, \bibinfo {author} {\bibfnamefont {K.}~\bibnamefont {{Barmak}}}, \bibinfo {author} {\bibfnamefont {J.}~\bibnamefont {{Hone}}}, \bibinfo {author} {\bibfnamefont {A.~J.}\ \bibnamefont {{Millis}}}, \bibinfo {author} {\bibfnamefont {A.~N.}\ \bibnamefont {{Pasupathy}}}, \ and\ \bibinfo {author} {\bibfnamefont {C.~R.}\ \bibnamefont {{Dean}}},\ }\bibfield  {title} {\enquote {\bibinfo {title} {{Superconductivity in twisted bilayer WSe$_2$}},}\ }\href {\doibase 10.48550/arXiv.2406.03418}
  {\bibfield  {journal} {\bibinfo  {journal} {arXiv e-prints}\ ,\ \bibinfo {eid} {arXiv:2406.03418}} (\bibinfo {year} {2024})},\ \Eprint {http://arxiv.org/abs/2406.03418} {arXiv:2406.03418 [cond-mat.mes-hall]} \BibitemShut {NoStop}%
\bibitem [{\citenamefont {Wang}\ \emph {et~al.}(2020)\citenamefont {Wang}, \citenamefont {Shih}, \citenamefont {Ghiotto}, \citenamefont {Xian}, \citenamefont {Rhodes}, \citenamefont {Tan}, \citenamefont {Claassen}, \citenamefont {Kennes}, \citenamefont {Bai}, \citenamefont {Kim}, \citenamefont {Watanabe}, \citenamefont {Taniguchi}, \citenamefont {Zhu}, \citenamefont {Hone}, \citenamefont {Rubio}, \citenamefont {Pasupathy},\ and\ \citenamefont {Dean}}]{Dean20}%
  \BibitemOpen
  \bibfield  {author} {\bibinfo {author} {\bibfnamefont {L.}~\bibnamefont {Wang}}, \bibinfo {author} {\bibfnamefont {E.-M.}\ \bibnamefont {Shih}}, \bibinfo {author} {\bibfnamefont {A.}~\bibnamefont {Ghiotto}}, \bibinfo {author} {\bibfnamefont {L.}~\bibnamefont {Xian}}, \bibinfo {author} {\bibfnamefont {D.~A.}\ \bibnamefont {Rhodes}}, \bibinfo {author} {\bibfnamefont {C.}~\bibnamefont {Tan}}, \bibinfo {author} {\bibfnamefont {M.}~\bibnamefont {Claassen}}, \bibinfo {author} {\bibfnamefont {D.~M.}\ \bibnamefont {Kennes}}, \bibinfo {author} {\bibfnamefont {Y.}~\bibnamefont {Bai}}, \bibinfo {author} {\bibfnamefont {B.}~\bibnamefont {Kim}}, \bibinfo {author} {\bibfnamefont {K.}~\bibnamefont {Watanabe}}, \bibinfo {author} {\bibfnamefont {T.}~\bibnamefont {Taniguchi}}, \bibinfo {author} {\bibfnamefont {X.}~\bibnamefont {Zhu}}, \bibinfo {author} {\bibfnamefont {J.}~\bibnamefont {Hone}}, \bibinfo {author} {\bibfnamefont {A.}~\bibnamefont {Rubio}}, \bibinfo {author} {\bibfnamefont {A.~N.}\ \bibnamefont {Pasupathy}}, \
  and\ \bibinfo {author} {\bibfnamefont {C.~R.}\ \bibnamefont {Dean}},\ }\bibfield  {title} {\enquote {\bibinfo {title} {Correlated electronic phases in twisted bilayer transition metal dichalcogenides},}\ }\href {\doibase 10.1038/s41563-020-0708-6} {\bibfield  {journal} {\bibinfo  {journal} {Nature Materials}\ }\textbf {\bibinfo {volume} {19}},\ \bibinfo {pages} {861} (\bibinfo {year} {2020})}\BibitemShut {NoStop}%
\bibitem [{\citenamefont {Kapitulnik}\ \emph {et~al.}(2019)\citenamefont {Kapitulnik}, \citenamefont {Kivelson},\ and\ \citenamefont {Spivak}}]{SKRMP}%
  \BibitemOpen
  \bibfield  {author} {\bibinfo {author} {\bibfnamefont {A.}~\bibnamefont {Kapitulnik}}, \bibinfo {author} {\bibfnamefont {S.~A.}\ \bibnamefont {Kivelson}}, \ and\ \bibinfo {author} {\bibfnamefont {B.}~\bibnamefont {Spivak}},\ }\bibfield  {title} {\enquote {\bibinfo {title} {Colloquium: Anomalous metals: Failed superconductors},}\ }\href {\doibase 10.1103/RevModPhys.91.011002} {\bibfield  {journal} {\bibinfo  {journal} {Rev. Mod. Phys.}\ }\textbf {\bibinfo {volume} {91}},\ \bibinfo {pages} {011002} (\bibinfo {year} {2019})}\BibitemShut {NoStop}%
\bibitem [{\citenamefont {Hofmann}\ \emph {et~al.}(2023)\citenamefont {Hofmann}, \citenamefont {Berg},\ and\ \citenamefont {Chowdhury}}]{DCPRL23}%
  \BibitemOpen
  \bibfield  {author} {\bibinfo {author} {\bibfnamefont {J.~S.}\ \bibnamefont {Hofmann}}, \bibinfo {author} {\bibfnamefont {E.}~\bibnamefont {Berg}}, \ and\ \bibinfo {author} {\bibfnamefont {D.}~\bibnamefont {Chowdhury}},\ }\bibfield  {title} {\enquote {\bibinfo {title} {Superconductivity, charge density wave, and supersolidity in flat bands with a tunable quantum metric},}\ }\href {\doibase 10.1103/PhysRevLett.130.226001} {\bibfield  {journal} {\bibinfo  {journal} {Phys. Rev. Lett.}\ }\textbf {\bibinfo {volume} {130}},\ \bibinfo {pages} {226001} (\bibinfo {year} {2023})}\BibitemShut {NoStop}%
\bibitem [{\citenamefont {{Schrade}}\ and\ \citenamefont {{Fu}}(2021)}]{constantin}%
  \BibitemOpen
  \bibfield  {author} {\bibinfo {author} {\bibfnamefont {C.}~\bibnamefont {{Schrade}}}\ and\ \bibinfo {author} {\bibfnamefont {L.}~\bibnamefont {{Fu}}},\ }\bibfield  {title} {\enquote {\bibinfo {title} {{Nematic, chiral and topological superconductivity in transition metal dichalcogenides}},}\ }\href {\doibase 10.48550/arXiv.2110.10172} {\bibfield  {journal} {\bibinfo  {journal} {arXiv e-prints}\ ,\ \bibinfo {eid} {arXiv:2110.10172}} (\bibinfo {year} {2021})},\ \Eprint {http://arxiv.org/abs/2110.10172} {arXiv:2110.10172 [cond-mat.supr-con]} \BibitemShut {NoStop}%
\bibitem [{\citenamefont {Hsu}\ \emph {et~al.}(2021)\citenamefont {Hsu}, \citenamefont {Wu},\ and\ \citenamefont {Das~Sarma}}]{SDS_VHS_RG}%
  \BibitemOpen
  \bibfield  {author} {\bibinfo {author} {\bibfnamefont {Y.-T.}\ \bibnamefont {Hsu}}, \bibinfo {author} {\bibfnamefont {F.}~\bibnamefont {Wu}}, \ and\ \bibinfo {author} {\bibfnamefont {S.}~\bibnamefont {Das~Sarma}},\ }\bibfield  {title} {\enquote {\bibinfo {title} {Spin-valley locked instabilities in moir\'e transition metal dichalcogenides with conventional and higher-order van hove singularities},}\ }\href {\doibase 10.1103/PhysRevB.104.195134} {\bibfield  {journal} {\bibinfo  {journal} {Phys. Rev. B}\ }\textbf {\bibinfo {volume} {104}},\ \bibinfo {pages} {195134} (\bibinfo {year} {2021})}\BibitemShut {NoStop}%
\bibitem [{\citenamefont {B\'elanger}\ \emph {et~al.}(2022)\citenamefont {B\'elanger}, \citenamefont {Fournier},\ and\ \citenamefont {S\'en\'echal}}]{Senechal_VHS_dmft}%
  \BibitemOpen
  \bibfield  {author} {\bibinfo {author} {\bibfnamefont {M.}~\bibnamefont {B\'elanger}}, \bibinfo {author} {\bibfnamefont {J.}~\bibnamefont {Fournier}}, \ and\ \bibinfo {author} {\bibfnamefont {D.}~\bibnamefont {S\'en\'echal}},\ }\bibfield  {title} {\enquote {\bibinfo {title} {Superconductivity in the twisted bilayer transition metal dichalcogenide ${\mathrm{wse}}_{2}$: A quantum cluster study},}\ }\href {\doibase 10.1103/PhysRevB.106.235135} {\bibfield  {journal} {\bibinfo  {journal} {Phys. Rev. B}\ }\textbf {\bibinfo {volume} {106}},\ \bibinfo {pages} {235135} (\bibinfo {year} {2022})}\BibitemShut {NoStop}%
\bibitem [{\citenamefont {Scherer}\ \emph {et~al.}(2022)\citenamefont {Scherer}, \citenamefont {Kennes},\ and\ \citenamefont {Classen}}]{Scherer_VHS_hetero}%
  \BibitemOpen
  \bibfield  {author} {\bibinfo {author} {\bibfnamefont {M.~M.}\ \bibnamefont {Scherer}}, \bibinfo {author} {\bibfnamefont {D.~M.}\ \bibnamefont {Kennes}}, \ and\ \bibinfo {author} {\bibfnamefont {L.}~\bibnamefont {Classen}},\ }\bibfield  {title} {\enquote {\bibinfo {title} {Chiral superconductivity with enhanced quantized hall responses in moiré transition metal dichalcogenides},}\ }\href {\doibase 10.1038/s41535-022-00504-z} {\bibfield  {journal} {\bibinfo  {journal} {npj Quantum Materials}\ }\textbf {\bibinfo {volume} {7}} (\bibinfo {year} {2022}),\ 10.1038/s41535-022-00504-z}\BibitemShut {NoStop}%
\bibitem [{\citenamefont {Klebl}\ \emph {et~al.}(2023)\citenamefont {Klebl}, \citenamefont {Fischer}, \citenamefont {Classen}, \citenamefont {Scherer},\ and\ \citenamefont {Kennes}}]{Kennes_VHS}%
  \BibitemOpen
  \bibfield  {author} {\bibinfo {author} {\bibfnamefont {L.}~\bibnamefont {Klebl}}, \bibinfo {author} {\bibfnamefont {A.}~\bibnamefont {Fischer}}, \bibinfo {author} {\bibfnamefont {L.}~\bibnamefont {Classen}}, \bibinfo {author} {\bibfnamefont {M.~M.}\ \bibnamefont {Scherer}}, \ and\ \bibinfo {author} {\bibfnamefont {D.~M.}\ \bibnamefont {Kennes}},\ }\bibfield  {title} {\enquote {\bibinfo {title} {Competition of density waves and superconductivity in twisted tungsten diselenide},}\ }\href {\doibase 10.1103/PhysRevResearch.5.L012034} {\bibfield  {journal} {\bibinfo  {journal} {Phys. Rev. Res.}\ }\textbf {\bibinfo {volume} {5}},\ \bibinfo {pages} {L012034} (\bibinfo {year} {2023})}\BibitemShut {NoStop}%
\bibitem [{\citenamefont {Wu}\ \emph {et~al.}(2023)\citenamefont {Wu}, \citenamefont {Wu},\ and\ \citenamefont {Yao}}]{HongYao_VHS}%
  \BibitemOpen
  \bibfield  {author} {\bibinfo {author} {\bibfnamefont {Y.-M.}\ \bibnamefont {Wu}}, \bibinfo {author} {\bibfnamefont {Z.}~\bibnamefont {Wu}}, \ and\ \bibinfo {author} {\bibfnamefont {H.}~\bibnamefont {Yao}},\ }\bibfield  {title} {\enquote {\bibinfo {title} {Pair-density-wave and chiral superconductivity in twisted bilayer transition metal dichalcogenides},}\ }\href {\doibase 10.1103/PhysRevLett.130.126001} {\bibfield  {journal} {\bibinfo  {journal} {Phys. Rev. Lett.}\ }\textbf {\bibinfo {volume} {130}},\ \bibinfo {pages} {126001} (\bibinfo {year} {2023})}\BibitemShut {NoStop}%
\bibitem [{\citenamefont {Zegrodnik}\ and\ \citenamefont {Biborski}(2023)}]{Biborski_VHS}%
  \BibitemOpen
  \bibfield  {author} {\bibinfo {author} {\bibfnamefont {M.}~\bibnamefont {Zegrodnik}}\ and\ \bibinfo {author} {\bibfnamefont {A.}~\bibnamefont {Biborski}},\ }\bibfield  {title} {\enquote {\bibinfo {title} {Mixed singlet-triplet superconducting state within the moir\'e $t\text{\ensuremath{-}}j\text{\ensuremath{-}}u$ model applied to twisted bilayer ${\mathrm{wse}}_{2}$},}\ }\href {\doibase 10.1103/PhysRevB.108.064506} {\bibfield  {journal} {\bibinfo  {journal} {Phys. Rev. B}\ }\textbf {\bibinfo {volume} {108}},\ \bibinfo {pages} {064506} (\bibinfo {year} {2023})}\BibitemShut {NoStop}%
\bibitem [{\citenamefont {{Akbar}}\ \emph {et~al.}(2024)\citenamefont {{Akbar}}, \citenamefont {{Biborski}}, \citenamefont {{Rademaker}},\ and\ \citenamefont {{Zegrodnik}}}]{Rademaker_VHS_strongcoupling}%
  \BibitemOpen
  \bibfield  {author} {\bibinfo {author} {\bibfnamefont {W.}~\bibnamefont {{Akbar}}}, \bibinfo {author} {\bibfnamefont {A.}~\bibnamefont {{Biborski}}}, \bibinfo {author} {\bibfnamefont {L.}~\bibnamefont {{Rademaker}}}, \ and\ \bibinfo {author} {\bibfnamefont {M.}~\bibnamefont {{Zegrodnik}}},\ }\bibfield  {title} {\enquote {\bibinfo {title} {{Topological superconductivity with mixed singlet-triplet pairing in moir{\'e} transition-metal-dichalcogenide bilayers}},}\ }\href {\doibase 10.48550/arXiv.2403.05903} {\bibfield  {journal} {\bibinfo  {journal} {arXiv e-prints}\ ,\ \bibinfo {eid} {arXiv:2403.05903}} (\bibinfo {year} {2024})},\ \Eprint {http://arxiv.org/abs/2403.05903} {arXiv:2403.05903 [cond-mat.supr-con]} \BibitemShut {NoStop}%
\bibitem [{\citenamefont {Venderley}\ and\ \citenamefont {Kim}(2019)}]{EAKim_doped_dmrg}%
  \BibitemOpen
  \bibfield  {author} {\bibinfo {author} {\bibfnamefont {J.}~\bibnamefont {Venderley}}\ and\ \bibinfo {author} {\bibfnamefont {E.-A.}\ \bibnamefont {Kim}},\ }\bibfield  {title} {\enquote {\bibinfo {title} {Density matrix renormalization group study of superconductivity in the triangular lattice hubbard model},}\ }\href {\doibase 10.1103/PhysRevB.100.060506} {\bibfield  {journal} {\bibinfo  {journal} {Phys. Rev. B}\ }\textbf {\bibinfo {volume} {100}},\ \bibinfo {pages} {060506} (\bibinfo {year} {2019})}\BibitemShut {NoStop}%
\bibitem [{\citenamefont {Slagle}\ and\ \citenamefont {Fu}(2020)}]{LiangFu_doped_trimer}%
  \BibitemOpen
  \bibfield  {author} {\bibinfo {author} {\bibfnamefont {K.}~\bibnamefont {Slagle}}\ and\ \bibinfo {author} {\bibfnamefont {L.}~\bibnamefont {Fu}},\ }\bibfield  {title} {\enquote {\bibinfo {title} {Charge transfer excitations, pair density waves, and superconductivity in moir\'e materials},}\ }\href {\doibase 10.1103/PhysRevB.102.235423} {\bibfield  {journal} {\bibinfo  {journal} {Phys. Rev. B}\ }\textbf {\bibinfo {volume} {102}},\ \bibinfo {pages} {235423} (\bibinfo {year} {2020})}\BibitemShut {NoStop}%
\bibitem [{\citenamefont {Chen}\ and\ \citenamefont {Sheng}(2023)}]{DNSheng_doped_dmrg}%
  \BibitemOpen
  \bibfield  {author} {\bibinfo {author} {\bibfnamefont {F.}~\bibnamefont {Chen}}\ and\ \bibinfo {author} {\bibfnamefont {D.~N.}\ \bibnamefont {Sheng}},\ }\bibfield  {title} {\enquote {\bibinfo {title} {Singlet, triplet, and pair density wave superconductivity in the doped triangular-lattice moir\'e system},}\ }\href {\doibase 10.1103/PhysRevB.108.L201110} {\bibfield  {journal} {\bibinfo  {journal} {Phys. Rev. B}\ }\textbf {\bibinfo {volume} {108}},\ \bibinfo {pages} {L201110} (\bibinfo {year} {2023})}\BibitemShut {NoStop}%
\bibitem [{\citenamefont {Cr\'epel}\ \emph {et~al.}(2023)\citenamefont {Cr\'epel}, \citenamefont {Guerci}, \citenamefont {Cano}, \citenamefont {Pixley},\ and\ \citenamefont {Millis}}]{Millis_doped_TSC}%
  \BibitemOpen
  \bibfield  {author} {\bibinfo {author} {\bibfnamefont {V.}~\bibnamefont {Cr\'epel}}, \bibinfo {author} {\bibfnamefont {D.}~\bibnamefont {Guerci}}, \bibinfo {author} {\bibfnamefont {J.}~\bibnamefont {Cano}}, \bibinfo {author} {\bibfnamefont {J.~H.}\ \bibnamefont {Pixley}}, \ and\ \bibinfo {author} {\bibfnamefont {A.}~\bibnamefont {Millis}},\ }\bibfield  {title} {\enquote {\bibinfo {title} {Topological superconductivity in doped magnetic moir\'e semiconductors},}\ }\href {\doibase 10.1103/PhysRevLett.131.056001} {\bibfield  {journal} {\bibinfo  {journal} {Phys. Rev. Lett.}\ }\textbf {\bibinfo {volume} {131}},\ \bibinfo {pages} {056001} (\bibinfo {year} {2023})}\BibitemShut {NoStop}%
\bibitem [{\citenamefont {Zhou}\ and\ \citenamefont {Zhang}(2023{\natexlab{a}})}]{Yahui_doped_parton}%
  \BibitemOpen
  \bibfield  {author} {\bibinfo {author} {\bibfnamefont {B.}~\bibnamefont {Zhou}}\ and\ \bibinfo {author} {\bibfnamefont {Y.-H.}\ \bibnamefont {Zhang}},\ }\bibfield  {title} {\enquote {\bibinfo {title} {Chiral and nodal superconductors in the $t\text{\ensuremath{-}}j$ model with valley contrasting flux on a triangular moir\'e lattice},}\ }\href {\doibase 10.1103/PhysRevB.108.155111} {\bibfield  {journal} {\bibinfo  {journal} {Phys. Rev. B}\ }\textbf {\bibinfo {volume} {108}},\ \bibinfo {pages} {155111} (\bibinfo {year} {2023}{\natexlab{a}})}\BibitemShut {NoStop}%
\bibitem [{\citenamefont {Xie}\ and\ \citenamefont {Law}(2023)}]{KTLaw_doped}%
  \BibitemOpen
  \bibfield  {author} {\bibinfo {author} {\bibfnamefont {Y.-M.}\ \bibnamefont {Xie}}\ and\ \bibinfo {author} {\bibfnamefont {K.~T.}\ \bibnamefont {Law}},\ }\bibfield  {title} {\enquote {\bibinfo {title} {Orbital fulde-ferrell pairing state in moir\'e ising superconductors},}\ }\href {\doibase 10.1103/PhysRevLett.131.016001} {\bibfield  {journal} {\bibinfo  {journal} {Phys. Rev. Lett.}\ }\textbf {\bibinfo {volume} {131}},\ \bibinfo {pages} {016001} (\bibinfo {year} {2023})}\BibitemShut {NoStop}%
\bibitem [{\citenamefont {Broholm}\ \emph {et~al.}(2020)\citenamefont {Broholm}, \citenamefont {Cava}, \citenamefont {Kivelson}, \citenamefont {Nocera}, \citenamefont {Norman},\ and\ \citenamefont {Senthil}}]{QSL}%
  \BibitemOpen
  \bibfield  {author} {\bibinfo {author} {\bibfnamefont {C.}~\bibnamefont {Broholm}}, \bibinfo {author} {\bibfnamefont {R.~J.}\ \bibnamefont {Cava}}, \bibinfo {author} {\bibfnamefont {S.~A.}\ \bibnamefont {Kivelson}}, \bibinfo {author} {\bibfnamefont {D.~G.}\ \bibnamefont {Nocera}}, \bibinfo {author} {\bibfnamefont {M.~R.}\ \bibnamefont {Norman}}, \ and\ \bibinfo {author} {\bibfnamefont {T.}~\bibnamefont {Senthil}},\ }\bibfield  {title} {\enquote {\bibinfo {title} {Quantum spin liquids},}\ }\href {https://science.sciencemag.org/content/367/6475/eaay0668} {\bibfield  {journal} {\bibinfo  {journal} {Science}\ }\textbf {\bibinfo {volume} {367}} (\bibinfo {year} {2020})}\BibitemShut {NoStop}%
\bibitem [{\citenamefont {Lee}\ \emph {et~al.}(2006)\citenamefont {Lee}, \citenamefont {Nagaosa},\ and\ \citenamefont {Wen}}]{LNW}%
  \BibitemOpen
  \bibfield  {author} {\bibinfo {author} {\bibfnamefont {P.~A.}\ \bibnamefont {Lee}}, \bibinfo {author} {\bibfnamefont {N.}~\bibnamefont {Nagaosa}}, \ and\ \bibinfo {author} {\bibfnamefont {X.-G.}\ \bibnamefont {Wen}},\ }\bibfield  {title} {\enquote {\bibinfo {title} {Doping a mott insulator: Physics of high-temperature superconductivity},}\ }\href {\doibase 10.1103/RevModPhys.78.17} {\bibfield  {journal} {\bibinfo  {journal} {Rev. Mod. Phys.}\ }\textbf {\bibinfo {volume} {78}},\ \bibinfo {pages} {17} (\bibinfo {year} {2006})}\BibitemShut {NoStop}%
\bibitem [{\citenamefont {Senthil}\ and\ \citenamefont {Fisher}(2000)}]{Senthil_Z2QSL_PRB}%
  \BibitemOpen
  \bibfield  {author} {\bibinfo {author} {\bibfnamefont {T.}~\bibnamefont {Senthil}}\ and\ \bibinfo {author} {\bibfnamefont {M.~P.~A.}\ \bibnamefont {Fisher}},\ }\bibfield  {title} {\enquote {\bibinfo {title} {${Z}_{2}$ gauge theory of electron fractionalization in strongly correlated systems},}\ }\href {\doibase 10.1103/PhysRevB.62.7850} {\bibfield  {journal} {\bibinfo  {journal} {Phys. Rev. B}\ }\textbf {\bibinfo {volume} {62}},\ \bibinfo {pages} {7850} (\bibinfo {year} {2000})}\BibitemShut {NoStop}%
\bibitem [{\citenamefont {{Cr\'epel}}\ and\ \citenamefont {{Millis}}(2024)}]{valentin24}%
  \BibitemOpen
  \bibfield  {author} {\bibinfo {author} {\bibfnamefont {V.}~\bibnamefont {{Cr\'epel}}}\ and\ \bibinfo {author} {\bibfnamefont {A.}~\bibnamefont {{Millis}}},\ }\bibfield  {title} {\enquote {\bibinfo {title} {{Bridging the small and large in twisted transition metal dicalcogenide homobilayers: a tight binding model capturing orbital interference and topology across a wide range of twist angles}},}\ }\href {\doibase 10.48550/arXiv.2403.15546} {\bibfield  {journal} {\bibinfo  {journal} {arXiv e-prints}\ ,\ \bibinfo {eid} {arXiv:2403.15546}} (\bibinfo {year} {2024})},\ \Eprint {http://arxiv.org/abs/2403.15546} {arXiv:2403.15546 [cond-mat.str-el]} \BibitemShut {NoStop}%
\bibitem [{\citenamefont {Wu}\ \emph {et~al.}(2019)\citenamefont {Wu}, \citenamefont {Lovorn}, \citenamefont {Tutuc}, \citenamefont {Martin},\ and\ \citenamefont {MacDonald}}]{Wu_twisted_prl}%
  \BibitemOpen
  \bibfield  {author} {\bibinfo {author} {\bibfnamefont {F.}~\bibnamefont {Wu}}, \bibinfo {author} {\bibfnamefont {T.}~\bibnamefont {Lovorn}}, \bibinfo {author} {\bibfnamefont {E.}~\bibnamefont {Tutuc}}, \bibinfo {author} {\bibfnamefont {I.}~\bibnamefont {Martin}}, \ and\ \bibinfo {author} {\bibfnamefont {A.~H.}\ \bibnamefont {MacDonald}},\ }\bibfield  {title} {\enquote {\bibinfo {title} {Topological insulators in twisted transition metal dichalcogenide homobilayers},}\ }\href {\doibase 10.1103/PhysRevLett.122.086402} {\bibfield  {journal} {\bibinfo  {journal} {Phys. Rev. Lett.}\ }\textbf {\bibinfo {volume} {122}},\ \bibinfo {pages} {086402} (\bibinfo {year} {2019})}\BibitemShut {NoStop}%
\bibitem [{\citenamefont {Devakul}\ \emph {et~al.}(2021)\citenamefont {Devakul}, \citenamefont {Cr{\'e}pel}, \citenamefont {Zhang},\ and\ \citenamefont {Fu}}]{Devakul_magic2021}%
  \BibitemOpen
  \bibfield  {author} {\bibinfo {author} {\bibfnamefont {T.}~\bibnamefont {Devakul}}, \bibinfo {author} {\bibfnamefont {V.}~\bibnamefont {Cr{\'e}pel}}, \bibinfo {author} {\bibfnamefont {Y.}~\bibnamefont {Zhang}}, \ and\ \bibinfo {author} {\bibfnamefont {L.}~\bibnamefont {Fu}},\ }\bibfield  {title} {\enquote {\bibinfo {title} {Magic in twisted transition metal dichalcogenide bilayers},}\ }\href {\doibase 10.1038/s41467-021-27042-9} {\bibfield  {journal} {\bibinfo  {journal} {Nature Communications}\ }\textbf {\bibinfo {volume} {12}},\ \bibinfo {pages} {6730} (\bibinfo {year} {2021})}\BibitemShut {NoStop}%
\bibitem [{\citenamefont {Xiao}\ \emph {et~al.}(2012)\citenamefont {Xiao}, \citenamefont {Liu}, \citenamefont {Feng}, \citenamefont {Xu},\ and\ \citenamefont {Yao}}]{Xiao_SOC_PRL}%
  \BibitemOpen
  \bibfield  {author} {\bibinfo {author} {\bibfnamefont {D.}~\bibnamefont {Xiao}}, \bibinfo {author} {\bibfnamefont {G.-B.}\ \bibnamefont {Liu}}, \bibinfo {author} {\bibfnamefont {W.}~\bibnamefont {Feng}}, \bibinfo {author} {\bibfnamefont {X.}~\bibnamefont {Xu}}, \ and\ \bibinfo {author} {\bibfnamefont {W.}~\bibnamefont {Yao}},\ }\bibfield  {title} {\enquote {\bibinfo {title} {Coupled spin and valley physics in monolayers of ${\mathrm{mos}}_{2}$ and other group-vi dichalcogenides},}\ }\href {\doibase 10.1103/PhysRevLett.108.196802} {\bibfield  {journal} {\bibinfo  {journal} {Phys. Rev. Lett.}\ }\textbf {\bibinfo {volume} {108}},\ \bibinfo {pages} {196802} (\bibinfo {year} {2012})}\BibitemShut {NoStop}%
\bibitem [{\citenamefont {Fang}\ \emph {et~al.}(2012)\citenamefont {Fang}, \citenamefont {Gilbert},\ and\ \citenamefont {Bernevig}}]{fang_invariants_2012}%
  \BibitemOpen
  \bibfield  {author} {\bibinfo {author} {\bibfnamefont {C.}~\bibnamefont {Fang}}, \bibinfo {author} {\bibfnamefont {M.~J.}\ \bibnamefont {Gilbert}}, \ and\ \bibinfo {author} {\bibfnamefont {B.~A.}\ \bibnamefont {Bernevig}},\ }\bibfield  {title} {\enquote {\bibinfo {title} {Bulk topological invariants in noninteracting point group symmetric insulators},}\ }\href {\doibase 10.1103/PhysRevB.86.115112} {\bibfield  {journal} {\bibinfo  {journal} {Phys. Rev. B}\ }\textbf {\bibinfo {volume} {86}},\ \bibinfo {pages} {115112} (\bibinfo {year} {2012})}\BibitemShut {NoStop}%
\bibitem [{\citenamefont {Yu}\ \emph {et~al.}(2019)\citenamefont {Yu}, \citenamefont {Chen},\ and\ \citenamefont {Yao}}]{3orb_Yu_NSR19}%
  \BibitemOpen
  \bibfield  {author} {\bibinfo {author} {\bibfnamefont {H.}~\bibnamefont {Yu}}, \bibinfo {author} {\bibfnamefont {M.}~\bibnamefont {Chen}}, \ and\ \bibinfo {author} {\bibfnamefont {W.}~\bibnamefont {Yao}},\ }\bibfield  {title} {\enquote {\bibinfo {title} {Giant magnetic field from moiré induced berry phase in homobilayer semiconductors},}\ }\href {\doibase 10.1093/nsr/nwz117} {\bibfield  {journal} {\bibinfo  {journal} {National Science Review}\ }\textbf {\bibinfo {volume} {7}},\ \bibinfo {pages} {12–20} (\bibinfo {year} {2019})}\BibitemShut {NoStop}%
\bibitem [{\citenamefont {Qiu}\ \emph {et~al.}(2023)\citenamefont {Qiu}, \citenamefont {Li}, \citenamefont {Luo},\ and\ \citenamefont {Wu}}]{3orb_Wu_PRX24}%
  \BibitemOpen
  \bibfield  {author} {\bibinfo {author} {\bibfnamefont {W.-X.}\ \bibnamefont {Qiu}}, \bibinfo {author} {\bibfnamefont {B.}~\bibnamefont {Li}}, \bibinfo {author} {\bibfnamefont {X.-J.}\ \bibnamefont {Luo}}, \ and\ \bibinfo {author} {\bibfnamefont {F.}~\bibnamefont {Wu}},\ }\bibfield  {title} {\enquote {\bibinfo {title} {Interaction-driven topological phase diagram of twisted bilayer ${\mathrm{mote}}_{2}$},}\ }\href {\doibase 10.1103/PhysRevX.13.041026} {\bibfield  {journal} {\bibinfo  {journal} {Phys. Rev. X}\ }\textbf {\bibinfo {volume} {13}},\ \bibinfo {pages} {041026} (\bibinfo {year} {2023})}\BibitemShut {NoStop}%
\bibitem [{\citenamefont {Xu}\ \emph {et~al.}(2024)\citenamefont {Xu}, \citenamefont {Li}, \citenamefont {Xu}, \citenamefont {Bi},\ and\ \citenamefont {Zhang}}]{3orb_Xu_PNAS24}%
  \BibitemOpen
  \bibfield  {author} {\bibinfo {author} {\bibfnamefont {C.}~\bibnamefont {Xu}}, \bibinfo {author} {\bibfnamefont {J.}~\bibnamefont {Li}}, \bibinfo {author} {\bibfnamefont {Y.}~\bibnamefont {Xu}}, \bibinfo {author} {\bibfnamefont {Z.}~\bibnamefont {Bi}}, \ and\ \bibinfo {author} {\bibfnamefont {Y.}~\bibnamefont {Zhang}},\ }\bibfield  {title} {\enquote {\bibinfo {title} {Maximally localized wannier functions, interaction models, and fractional quantum anomalous hall effect in twisted bilayer mote 2},}\ }\href {\doibase 10.1073/pnas.2316749121} {\bibfield  {journal} {\bibinfo  {journal} {Proceedings of the National Academy of Sciences}\ }\textbf {\bibinfo {volume} {121}} (\bibinfo {year} {2024}),\ 10.1073/pnas.2316749121}\BibitemShut {NoStop}%
\bibitem [{\citenamefont {Pan}\ \emph {et~al.}(2020)\citenamefont {Pan}, \citenamefont {Wu},\ and\ \citenamefont {Das~Sarma}}]{Haining_band_topo_twse_2020}%
  \BibitemOpen
  \bibfield  {author} {\bibinfo {author} {\bibfnamefont {H.}~\bibnamefont {Pan}}, \bibinfo {author} {\bibfnamefont {F.}~\bibnamefont {Wu}}, \ and\ \bibinfo {author} {\bibfnamefont {S.}~\bibnamefont {Das~Sarma}},\ }\bibfield  {title} {\enquote {\bibinfo {title} {Band topology, hubbard model, heisenberg model, and dzyaloshinskii-moriya interaction in twisted bilayer ${\mathrm{wse}}_{2}$},}\ }\href {\doibase 10.1103/PhysRevResearch.2.033087} {\bibfield  {journal} {\bibinfo  {journal} {Phys. Rev. Res.}\ }\textbf {\bibinfo {volume} {2}},\ \bibinfo {pages} {033087} (\bibinfo {year} {2020})}\BibitemShut {NoStop}%
\bibitem [{\citenamefont {{Zhang}}\ \emph {et~al.}(2024)\citenamefont {{Zhang}}, \citenamefont {{Morales-Dur{\'a}n}}, \citenamefont {{Li}}, \citenamefont {{Yao}}, \citenamefont {{Su}}, \citenamefont {{Lin}}, \citenamefont {{Dong}}, \citenamefont {{Kim}}, \citenamefont {{Robinson}}, \citenamefont {{Macdonald}},\ and\ \citenamefont {{Shih}}}]{AustinExpt}%
  \BibitemOpen
  \bibfield  {author} {\bibinfo {author} {\bibfnamefont {F.}~\bibnamefont {{Zhang}}}, \bibinfo {author} {\bibfnamefont {N.}~\bibnamefont {{Morales-Dur{\'a}n}}}, \bibinfo {author} {\bibfnamefont {Y.}~\bibnamefont {{Li}}}, \bibinfo {author} {\bibfnamefont {W.}~\bibnamefont {{Yao}}}, \bibinfo {author} {\bibfnamefont {J.-J.}\ \bibnamefont {{Su}}}, \bibinfo {author} {\bibfnamefont {Y.-C.}\ \bibnamefont {{Lin}}}, \bibinfo {author} {\bibfnamefont {C.}~\bibnamefont {{Dong}}}, \bibinfo {author} {\bibfnamefont {H.}~\bibnamefont {{Kim}}}, \bibinfo {author} {\bibfnamefont {J.~A.}\ \bibnamefont {{Robinson}}}, \bibinfo {author} {\bibfnamefont {A.~H.}\ \bibnamefont {{Macdonald}}}, \ and\ \bibinfo {author} {\bibfnamefont {C.-K.}\ \bibnamefont {{Shih}}},\ }\bibfield  {title} {\enquote {\bibinfo {title} {{Direct observation of layer skyrmions in twisted WSe2 bilayers}},}\ }\href {\doibase 10.48550/arXiv.2406.20036} {\bibfield  {journal} {\bibinfo  {journal} {arXiv e-prints}\ ,\ \bibinfo {eid} {arXiv:2406.20036}} (\bibinfo
  {year} {2024})},\ \Eprint {http://arxiv.org/abs/2406.20036} {arXiv:2406.20036 [cond-mat.mes-hall]} \BibitemShut {NoStop}%
\bibitem [{\citenamefont {Foutty}\ \emph {et~al.}(2024)\citenamefont {Foutty}, \citenamefont {Kometter}, \citenamefont {Devakul}, \citenamefont {Reddy}, \citenamefont {Watanabe}, \citenamefont {Taniguchi}, \citenamefont {Fu},\ and\ \citenamefont {Feldman}}]{Foutty_2024}%
  \BibitemOpen
  \bibfield  {author} {\bibinfo {author} {\bibfnamefont {B.~A.}\ \bibnamefont {Foutty}}, \bibinfo {author} {\bibfnamefont {C.~R.}\ \bibnamefont {Kometter}}, \bibinfo {author} {\bibfnamefont {T.}~\bibnamefont {Devakul}}, \bibinfo {author} {\bibfnamefont {A.~P.}\ \bibnamefont {Reddy}}, \bibinfo {author} {\bibfnamefont {K.}~\bibnamefont {Watanabe}}, \bibinfo {author} {\bibfnamefont {T.}~\bibnamefont {Taniguchi}}, \bibinfo {author} {\bibfnamefont {L.}~\bibnamefont {Fu}}, \ and\ \bibinfo {author} {\bibfnamefont {B.~E.}\ \bibnamefont {Feldman}},\ }\bibfield  {title} {\enquote {\bibinfo {title} {Mapping twist-tuned multiband topology in bilayer wse 2},}\ }\href {\doibase 10.1126/science.adi4728} {\bibfield  {journal} {\bibinfo  {journal} {Science}\ }\textbf {\bibinfo {volume} {384}},\ \bibinfo {pages} {343–347} (\bibinfo {year} {2024})}\BibitemShut {NoStop}%
\bibitem [{\citenamefont {Zhou}\ and\ \citenamefont {Zhang}(2023{\natexlab{b}})}]{yahui_chiral_2023}%
  \BibitemOpen
  \bibfield  {author} {\bibinfo {author} {\bibfnamefont {B.}~\bibnamefont {Zhou}}\ and\ \bibinfo {author} {\bibfnamefont {Y.-H.}\ \bibnamefont {Zhang}},\ }\bibfield  {title} {\enquote {\bibinfo {title} {Chiral and nodal superconductors in the $t\text{\ensuremath{-}}j$ model with valley contrasting flux on a triangular moir\'e lattice},}\ }\href {\doibase 10.1103/PhysRevB.108.155111} {\bibfield  {journal} {\bibinfo  {journal} {Phys. Rev. B}\ }\textbf {\bibinfo {volume} {108}},\ \bibinfo {pages} {155111} (\bibinfo {year} {2023}{\natexlab{b}})}\BibitemShut {NoStop}%
\bibitem [{\citenamefont {Senthil}\ and\ \citenamefont {Fisher}(2001)}]{Senthil_SC_spinon}%
  \BibitemOpen
  \bibfield  {author} {\bibinfo {author} {\bibfnamefont {T.}~\bibnamefont {Senthil}}\ and\ \bibinfo {author} {\bibfnamefont {M.~P.~A.}\ \bibnamefont {Fisher}},\ }\bibfield  {title} {\enquote {\bibinfo {title} {Fractionalization, topological order, and cuprate superconductivity},}\ }\href {\doibase 10.1103/PhysRevB.63.134521} {\bibfield  {journal} {\bibinfo  {journal} {Phys. Rev. B}\ }\textbf {\bibinfo {volume} {63}},\ \bibinfo {pages} {134521} (\bibinfo {year} {2001})}\BibitemShut {NoStop}%
\bibitem [{\citenamefont {Grover}\ \emph {et~al.}(2010)\citenamefont {Grover}, \citenamefont {Trivedi}, \citenamefont {Senthil},\ and\ \citenamefont {Lee}}]{Grover_SC_spinon}%
  \BibitemOpen
  \bibfield  {author} {\bibinfo {author} {\bibfnamefont {T.}~\bibnamefont {Grover}}, \bibinfo {author} {\bibfnamefont {N.}~\bibnamefont {Trivedi}}, \bibinfo {author} {\bibfnamefont {T.}~\bibnamefont {Senthil}}, \ and\ \bibinfo {author} {\bibfnamefont {P.~A.}\ \bibnamefont {Lee}},\ }\bibfield  {title} {\enquote {\bibinfo {title} {Weak mott insulators on the triangular lattice: Possibility of a gapless nematic quantum spin liquid},}\ }\href {\doibase 10.1103/PhysRevB.81.245121} {\bibfield  {journal} {\bibinfo  {journal} {Phys. Rev. B}\ }\textbf {\bibinfo {volume} {81}},\ \bibinfo {pages} {245121} (\bibinfo {year} {2010})}\BibitemShut {NoStop}%
\bibitem [{\citenamefont {Chowdhury}\ \emph {et~al.}(2018)\citenamefont {Chowdhury}, \citenamefont {Sodemann},\ and\ \citenamefont {Senthil}}]{DCNC}%
  \BibitemOpen
  \bibfield  {author} {\bibinfo {author} {\bibfnamefont {D.}~\bibnamefont {Chowdhury}}, \bibinfo {author} {\bibfnamefont {I.}~\bibnamefont {Sodemann}}, \ and\ \bibinfo {author} {\bibfnamefont {T.}~\bibnamefont {Senthil}},\ }\bibfield  {title} {\enquote {\bibinfo {title} {Mixed-valence insulators with neutral fermi surfaces},}\ }\href {\doibase 10.1038/s41467-018-04163-2} {\bibfield  {journal} {\bibinfo  {journal} {Nature Communications}\ }\textbf {\bibinfo {volume} {9}},\ \bibinfo {pages} {1766} (\bibinfo {year} {2018})}\BibitemShut {NoStop}%
\bibitem [{\citenamefont {Mendez-Valderrama}\ \emph {et~al.}(2024)\citenamefont {Mendez-Valderrama}, \citenamefont {Kim},\ and\ \citenamefont {Chowdhury}}]{SKDC24}%
  \BibitemOpen
  \bibfield  {author} {\bibinfo {author} {\bibfnamefont {J.~F.}\ \bibnamefont {Mendez-Valderrama}}, \bibinfo {author} {\bibfnamefont {S.}~\bibnamefont {Kim}}, \ and\ \bibinfo {author} {\bibfnamefont {D.}~\bibnamefont {Chowdhury}},\ }\bibfield  {title} {\enquote {\bibinfo {title} {Correlated topological mixed-valence insulators in moir\'e heterobilayers},}\ }\href {\doibase 10.1103/PhysRevB.110.L201105} {\bibfield  {journal} {\bibinfo  {journal} {Phys. Rev. B}\ }\textbf {\bibinfo {volume} {110}},\ \bibinfo {pages} {L201105} (\bibinfo {year} {2024})}\BibitemShut {NoStop}%
\bibitem [{\citenamefont {Metlitski}\ \emph {et~al.}(2015)\citenamefont {Metlitski}, \citenamefont {Mross}, \citenamefont {Sachdev},\ and\ \citenamefont {Senthil}}]{metlitski15}%
  \BibitemOpen
  \bibfield  {author} {\bibinfo {author} {\bibfnamefont {M.~A.}\ \bibnamefont {Metlitski}}, \bibinfo {author} {\bibfnamefont {D.~F.}\ \bibnamefont {Mross}}, \bibinfo {author} {\bibfnamefont {S.}~\bibnamefont {Sachdev}}, \ and\ \bibinfo {author} {\bibfnamefont {T.}~\bibnamefont {Senthil}},\ }\bibfield  {title} {\enquote {\bibinfo {title} {Cooper pairing in non-fermi liquids},}\ }\href {\doibase 10.1103/PhysRevB.91.115111} {\bibfield  {journal} {\bibinfo  {journal} {Phys. Rev. B}\ }\textbf {\bibinfo {volume} {91}},\ \bibinfo {pages} {115111} (\bibinfo {year} {2015})}\BibitemShut {NoStop}%
\bibitem [{\citenamefont {Li}\ \emph {et~al.}(2021)\citenamefont {Li}, \citenamefont {Jiang}, \citenamefont {Li}, \citenamefont {Zhang}, \citenamefont {Kang}, \citenamefont {Zhu}, \citenamefont {Watanabe}, \citenamefont {Taniguchi}, \citenamefont {Chowdhury}, \citenamefont {Fu} \emph {et~al.}}]{li2021continuous}%
  \BibitemOpen
  \bibfield  {author} {\bibinfo {author} {\bibfnamefont {T.}~\bibnamefont {Li}}, \bibinfo {author} {\bibfnamefont {S.}~\bibnamefont {Jiang}}, \bibinfo {author} {\bibfnamefont {L.}~\bibnamefont {Li}}, \bibinfo {author} {\bibfnamefont {Y.}~\bibnamefont {Zhang}}, \bibinfo {author} {\bibfnamefont {K.}~\bibnamefont {Kang}}, \bibinfo {author} {\bibfnamefont {J.}~\bibnamefont {Zhu}}, \bibinfo {author} {\bibfnamefont {K.}~\bibnamefont {Watanabe}}, \bibinfo {author} {\bibfnamefont {T.}~\bibnamefont {Taniguchi}}, \bibinfo {author} {\bibfnamefont {D.}~\bibnamefont {Chowdhury}}, \bibinfo {author} {\bibfnamefont {L.}~\bibnamefont {Fu}},  \emph {et~al.},\ }\bibfield  {title} {\enquote {\bibinfo {title} {Continuous mott transition in semiconductor moir{\'e} superlattices},}\ }\href {\doibase https://doi.org/10.1038/s41586-021-03853-0} {\bibfield  {journal} {\bibinfo  {journal} {Nature}\ }\textbf {\bibinfo {volume} {597}},\ \bibinfo {pages} {350} (\bibinfo {year} {2021})}\BibitemShut {NoStop}%
\bibitem [{\citenamefont {Sachdev}(2008)}]{ssrev}%
  \BibitemOpen
  \bibfield  {author} {\bibinfo {author} {\bibfnamefont {S.}~\bibnamefont {Sachdev}},\ }\bibfield  {title} {\enquote {\bibinfo {title} {Quantum magnetism and criticality},}\ }\href {\doibase 10.1038/nphys894} {\bibfield  {journal} {\bibinfo  {journal} {Nature Physics}\ }\textbf {\bibinfo {volume} {4}},\ \bibinfo {pages} {173} (\bibinfo {year} {2008})}\BibitemShut {NoStop}%
\bibitem [{\citenamefont {Senthil}(2008)}]{senthil_08}%
  \BibitemOpen
  \bibfield  {author} {\bibinfo {author} {\bibfnamefont {T.}~\bibnamefont {Senthil}},\ }\bibfield  {title} {\enquote {\bibinfo {title} {Theory of a continuous mott transition in two dimensions},}\ }\href {\doibase 10.1103/PhysRevB.78.045109} {\bibfield  {journal} {\bibinfo  {journal} {Phys. Rev. B}\ }\textbf {\bibinfo {volume} {78}},\ \bibinfo {pages} {045109} (\bibinfo {year} {2008})}\BibitemShut {NoStop}%
\bibitem [{\citenamefont {Szasz}\ \emph {et~al.}(2020)\citenamefont {Szasz}, \citenamefont {Motruk}, \citenamefont {Zaletel},\ and\ \citenamefont {Moore}}]{Szasz_CSL_PRX}%
  \BibitemOpen
  \bibfield  {author} {\bibinfo {author} {\bibfnamefont {A.}~\bibnamefont {Szasz}}, \bibinfo {author} {\bibfnamefont {J.}~\bibnamefont {Motruk}}, \bibinfo {author} {\bibfnamefont {M.~P.}\ \bibnamefont {Zaletel}}, \ and\ \bibinfo {author} {\bibfnamefont {J.~E.}\ \bibnamefont {Moore}},\ }\bibfield  {title} {\enquote {\bibinfo {title} {Chiral spin liquid phase of the triangular lattice hubbard model: A density matrix renormalization group study},}\ }\href {\doibase 10.1103/PhysRevX.10.021042} {\bibfield  {journal} {\bibinfo  {journal} {Phys. Rev. X}\ }\textbf {\bibinfo {volume} {10}},\ \bibinfo {pages} {021042} (\bibinfo {year} {2020})}\BibitemShut {NoStop}%
\bibitem [{\citenamefont {{Myerson-Jain}}\ and\ \citenamefont {{Xu}}(2024)}]{Cenke_DQCP}%
  \BibitemOpen
  \bibfield  {author} {\bibinfo {author} {\bibfnamefont {N.}~\bibnamefont {{Myerson-Jain}}}\ and\ \bibinfo {author} {\bibfnamefont {C.}~\bibnamefont {{Xu}}},\ }\bibfield  {title} {\enquote {\bibinfo {title} {{Superconductor-Insulator Transition in the TMD moir{\'e} systems and the Deconfined Quantum Critical Point}},}\ }\href {\doibase 10.48550/arXiv.2406.12971} {\bibfield  {journal} {\bibinfo  {journal} {arXiv e-prints}\ ,\ \bibinfo {eid} {arXiv:2406.12971}} (\bibinfo {year} {2024})},\ \Eprint {http://arxiv.org/abs/2406.12971} {arXiv:2406.12971 [cond-mat.str-el]} \BibitemShut {NoStop}%
\bibitem [{\citenamefont {Fisher}\ \emph {et~al.}(1989)\citenamefont {Fisher}, \citenamefont {Weichman}, \citenamefont {Grinstein},\ and\ \citenamefont {Fisher}}]{SFMI}%
  \BibitemOpen
  \bibfield  {author} {\bibinfo {author} {\bibfnamefont {M.~P.~A.}\ \bibnamefont {Fisher}}, \bibinfo {author} {\bibfnamefont {P.~B.}\ \bibnamefont {Weichman}}, \bibinfo {author} {\bibfnamefont {G.}~\bibnamefont {Grinstein}}, \ and\ \bibinfo {author} {\bibfnamefont {D.~S.}\ \bibnamefont {Fisher}},\ }\bibfield  {title} {\enquote {\bibinfo {title} {Boson localization and the superfluid-insulator transition},}\ }\href {\doibase 10.1103/PhysRevB.40.546} {\bibfield  {journal} {\bibinfo  {journal} {Phys. Rev. B}\ }\textbf {\bibinfo {volume} {40}},\ \bibinfo {pages} {546} (\bibinfo {year} {1989})}\BibitemShut {NoStop}%
\bibitem [{\citenamefont {Senthil}\ and\ \citenamefont {Lee}(2009)}]{TSPAL}%
  \BibitemOpen
  \bibfield  {author} {\bibinfo {author} {\bibfnamefont {T.}~\bibnamefont {Senthil}}\ and\ \bibinfo {author} {\bibfnamefont {P.~A.}\ \bibnamefont {Lee}},\ }\bibfield  {title} {\enquote {\bibinfo {title} {Coherence and pairing in a doped mott insulator: Application to the cuprates},}\ }\href {\doibase 10.1103/PhysRevLett.103.076402} {\bibfield  {journal} {\bibinfo  {journal} {Phys. Rev. Lett.}\ }\textbf {\bibinfo {volume} {103}},\ \bibinfo {pages} {076402} (\bibinfo {year} {2009})}\BibitemShut {NoStop}%
\bibitem [{\citenamefont {Cr\'epel}\ and\ \citenamefont {Millis}(2024)}]{Valentin_spinon_pairing}%
  \BibitemOpen
  \bibfield  {author} {\bibinfo {author} {\bibfnamefont {V.}~\bibnamefont {Cr\'epel}}\ and\ \bibinfo {author} {\bibfnamefont {A.}~\bibnamefont {Millis}},\ }\bibfield  {title} {\enquote {\bibinfo {title} {Spinon pairing induced by chiral in-plane exchange and the stabilization of odd-spin chern number spin liquid in twisted ${\mathrm{mote}}_{2}$},}\ }\href {\doibase 10.1103/PhysRevLett.133.146503} {\bibfield  {journal} {\bibinfo  {journal} {Phys. Rev. Lett.}\ }\textbf {\bibinfo {volume} {133}},\ \bibinfo {pages} {146503} (\bibinfo {year} {2024})}\BibitemShut {NoStop}%
\bibitem [{\citenamefont {Florens}\ and\ \citenamefont {Georges}(2004)}]{Florens_PRB_2004}%
  \BibitemOpen
  \bibfield  {author} {\bibinfo {author} {\bibfnamefont {S.}~\bibnamefont {Florens}}\ and\ \bibinfo {author} {\bibfnamefont {A.}~\bibnamefont {Georges}},\ }\bibfield  {title} {\enquote {\bibinfo {title} {Slave-rotor mean-field theories of strongly correlated systems and the mott transition in finite dimensions},}\ }\href {\doibase 10.1103/PhysRevB.70.035114} {\bibfield  {journal} {\bibinfo  {journal} {Phys. Rev. B}\ }\textbf {\bibinfo {volume} {70}},\ \bibinfo {pages} {035114} (\bibinfo {year} {2004})}\BibitemShut {NoStop}%
\bibitem [{\citenamefont {Zhao}\ and\ \citenamefont {Paramekanti}(2007)}]{Zhao_PRB_2007}%
  \BibitemOpen
  \bibfield  {author} {\bibinfo {author} {\bibfnamefont {E.}~\bibnamefont {Zhao}}\ and\ \bibinfo {author} {\bibfnamefont {A.}~\bibnamefont {Paramekanti}},\ }\bibfield  {title} {\enquote {\bibinfo {title} {Self-consistent slave rotor mean-field theory for strongly correlated systems},}\ }\href {\doibase 10.1103/PhysRevB.76.195101} {\bibfield  {journal} {\bibinfo  {journal} {Phys. Rev. B}\ }\textbf {\bibinfo {volume} {76}},\ \bibinfo {pages} {195101} (\bibinfo {year} {2007})}\BibitemShut {NoStop}%
\end{thebibliography}%
{\it Acknowledgements.-} We are indebted to Zhongdong Han, Kin-Fai Mak, Jie Shan and Yiyu Xia for numerous insightful discussions regarding their experimental results. This work is supported in part by a CAREER grant from the NSF to D.C. (DMR-2237522) and by a Sloan research fellowship from the Alfred P. Sloan foundation. 

\textit{Author contributions.-}  D.C. conceived and supervised the project. S.K., J.F.M.V. and X.W. performed the theoretical computations described in the paper. All authors contributed to interpretation of the results and writing of the manuscript. 

\textit{Competing Interests.-} The authors declare no competing interests.
\clearpage
\begin{widetext}
\renewcommand{\thefigure}{S\arabic{figure}}
\renewcommand{\figurename}{Supplemental Figure}
\setcounter{figure}{0}
\setcounter{section}{0}
\newcounter{suppfigure} 
\renewcommand{\thesuppfigure}{S\arabic{suppfigure}} 
\begin{widetext}
\begin{center}
    {\bf Supplementary material for ``Theory of Correlated Insulators and Superconductor at $\nu=1$ in Twisted WSe$_2$"}\\
    Sunghoon Kim$^*$, Juan Felipe Mendez-Valderrama$^*$,  Xuepeng Wang$^*$, Debanjan Chowdhury
\end{center}  

\section{Details of parton mean-field calculations}
\label{sec:SI_parton}

In this section, we provide details of our parton mean-field calculations. The parton representation for the electron at site $\vec{r}$ is given by $c_{\vec{r},\ell,\sigma}=b_{\vec{r},\ell}f_{\vec{r},\ell,\sigma}$, where $\ell$ and $\sigma$ denote the orbital and spin degrees of freedom, respectively. We impose the local constraints for the rotor charge and the spinon occupation, $\langle \sum_{\ell} n_{\vec{r},\ell}^\theta \rangle + \langle \sum_{\sigma,\ell} n^f_{\vec{r},\ell,\sigma}\rangle = \tn{const}$. The parton mean-field Hamiltonian can be written as
\beq \label{eq:mf_tot}
H&=&H^f +H^\theta + H^f_{\tn{ext}} +H^\theta_{\tn{ext}}, \nn \\
H^f&=&-\sum_{i\ne j,\sigma}t_{ij,\sigma}B_{ij}f^\dagger_{i\sigma}f_{j\sigma}-\mu^f \sum_{i}n_i^f  + J \sum'_{i\in T,j\in H} \bigg[e^{i\phi_{i,j}} f^{\dagger}_{i\uparrow}f^{\phantom\dagger}_{i\downarrow}f^{\dagger}_{j \downarrow}f^{\phantom\dagger}_{j \uparrow} + \tn{h.c.} \bigg] , \nn\\
H^\theta &=&-\sum_{i\ne j}\sum_\sigma t_{ij,\sigma} \chi_{ij,\sigma}b^\dagger_{i}b_{j} + \sum_i \frac{U_i}{2}(2-n_i^\theta)(1-n_i^\theta)+\sum_{ij}V_{ij}(2-n_i^\theta)(2-n_j^\theta)- \sum_i \mu_i^\theta n^\theta_i , \nn\\
H^f_{\tn{ext}} &=& (1-\alpha_{\theta}) E_z \sum_i (n_{i,\tn{XM}}^f - n_{i,\tn{MX}}^f) - (1-\beta_{\theta}) \delta \sum_{i}n^f_{i,\tn{MM}}, \nn\\
H^{\theta}_{\tn{ext}} &=& \alpha_{\theta} E_z \sum_{i} (n_{i,\tn{MX}}^{\theta} - n_{i,\tn{XM}}^{\theta}) - \beta_{\theta}\delta \sum_{i} (2-n^\theta_{i,\tn{MM}})
\eeq 
where $i$ is the combined index for the unit cell and orbital, $B_{ij}\equiv \langle b^\dagger_i b_j \rangle_\theta$ and $\chi_{ij,\sigma}\equiv \langle f_{i\sigma}^\dagger f_{j\sigma}\rangle_f$ are variational parameters. $\mu_{i}^{\theta,f}$ represent the bosonic and spinonic local chemical potentials introduced to impose the local constraints, respectively. The \textit{a priori} unknown bosonic layer polarization susceptibility is denoted $\alpha_\theta$ and $\beta_\theta$. The bosonic Hamiltonian is solved using a three-site cluster approximation, where each site corresponds to one of the three orbitals; see below for details of the cluster approximation. 

We perform self-consistent calculations in the following fashion. We restrict our attention to translationally-invariant solutions, while allowing for the rotational symmetry breaking of the spinon and boson correlators. We start by constructing initial $H^f$ with an ansatz for the spinonic correlator $\{\chi_{ij,\sigma}^{(0)}\}$ and the fermionic pairing $\{\Delta^{(0)}_{ij,\sigma\sigma'}\}$. Using the initial spinonic ground state, we compute $\{\chi_{ij,\sigma}^{(1)}\}$ and $\{\Delta^{(1)}_{ij,\sigma\sigma'}\}$, while imposing the filling constraints for the spinons. We then construct initial $H^\theta _{ijk}$'s for all mutually-connected 3-site clusters using $\{\chi_{ij,\sigma}^{(1)}\}$ and appropriate Lagrange multipliers $\{\mu_i^\theta\}$, which are chosen to satisfy the bosonic occupation constraints. We compute bosonic expectation values $\{B_{ij}^{(1)}\},\{\langle b_i \rangle^{(1)}\}$ using the initial $H^\theta _{ijk}$'s, where $\{\langle b_i \rangle^{(1)}\}$ is obtained by averaging over the results of all the cluster Hamiltonians involving the site $i$. We repeat this procedure to obtain a new set of variational parameters  
$\{\chi_{ij,\sigma}^{(2)}\},\{\Delta^{(2)}_{ij,\sigma\sigma'}\},\{B_{ij}^{(2)}\}$, and $\{\langle b_i \rangle^{(2)}\}$. The self-consistent calculations are performed until the correlators converge within a small threshold.

\subsubsection{Spinon mean-field Hamiltonian}
Let us start by defining the spinon pairing operators as
\begin{subequations}
\beq
\hat{\Phi}^{\alpha}_{S=0}(\vec{p})&\equiv&\bigg[ e^{i \alpha\pi/3}e^{i\vec{p}\cdot\vec{v}_{\alpha}} f_{\tau(\alpha),\downarrow}(-\vec{p}) f_{\tn{MM},\uparrow}(\p) +  e^{-i \alpha\pi/3}e^{-i\vec{p}\cdot\vec{v}_{\alpha}} f_{\tn{MM},\downarrow}(-\vec{p}) f_{\tau(\alpha),\uparrow}(\p)\bigg]\\
\hat{\Phi}^{\alpha}_{S=1}(\vec{p})&\equiv&\bigg[ e^{i \alpha\pi/3}e^{i\vec{p}\cdot\vec{v}_{\alpha}} f_{\tau(\alpha),\downarrow}(-\vec{p}) f_{\tn{MM},\uparrow}(\p) -  e^{-i \alpha\pi/3}e^{-i\vec{p}\cdot\vec{v}_{\alpha}} f_{\tn{MM},\downarrow}(-\vec{p}) f_{\tau(\alpha),\uparrow}(\p)\bigg],
\eeq
\end{subequations}
where $\alpha = 0,1,...,5$. Here, $\vec{v}_{2j} \equiv \vec{u}_{\tn{mod}(1-j,3)}$ for $j=0,1,2$ are defined as the vector pointing from MM-site to the nearest-neighbor MX-site; whereas $\vec{v}_{2j+1} \equiv -\vec{u}_{2-j}$ for $j=0,1,2$ are defined as the vector pointing from MM-site to the nearest-neighbor XM-site. The layer spinor $\tau(\alpha) \equiv \tn{mod}(\alpha, 2)$ is defined such that $\tau =0$ and $\tau=1$ denote XM-site and MX-site, respectively. The chiral-exchange interaction has the mean-field decomposition in the pairing channel as
\beq\label{spinon_pp}
\begin{aligned}
H^{\tn{pp}}_{\tn{chiral}}&=-J\sum_{\alpha}\bigg(\bigg[\sum_{\p}\hat{\Phi}^{\alpha}_{S=0}(\p)\bigg]^{\dagger}\bigg[\sum_{\k}\hat{\Phi}^{\alpha}_{S=0}(\k)\bigg] - \bigg[\sum_{\p}\hat{\Phi}^{\alpha}_{S=1}(\p)\bigg]^{\dagger}\bigg[\sum_{\k}\hat{\Phi}^{\alpha}_{S=1}(\k)\bigg]\bigg)\\
&\rightarrow-J\sum_{\alpha,\k}\bigg(\Delta^{\alpha\dagger}_{S=0} \hat{\Phi}^{\alpha}_{S=0}(\k) - \Delta^{\alpha\dagger}_{S=1} \hat{\Phi}^{\alpha}_{S=1}(\k) ~+~ \tn{h.c.} \bigg),
\end{aligned}
\eeq
where $\Delta^{\alpha}_{S}\equiv \avg{ \sum_{\k}\hat{\Phi}^{\alpha}_{\tn{S}}(\k)}$ for $S=0,1$. It can be easily seen that under C$_{3}$ rotation, $\Delta^{\alpha}_{S}\rightarrow \Delta^{\alpha+2}_{S}$, allowing us to define the angular momentum $L_z$ of the pairing mean-field as
\beq
\Delta^{\tau}_{L_z, S} \equiv  \sum^{2}_{j=0} e^{i 2jL_z\pi/3}\Delta^{2j+\tau}_{S},
\eeq
which corresponds to the pairing order parameters shown in Fig.\ref{fig4_spinon}c in the main text.

Similarly, the mean-field decomposition in the particle-hole channel is given by
\beq\label{spinon_ph}
H^{\tn{ph}}_{\tn{chiral}} \rightarrow J \sum_{\alpha,\sigma}\bigg[ \chi_{\Bar{\sigma}}^{\alpha} e^{i\alpha\pi/3} e^{i\vec{p}\cdot\vec{v}_{\alpha}} f^{\dagger}_{\tau(\alpha),\sigma}(\p) f_{\tn{MM},\sigma} (\p) + ~\tn{h.c.} \bigg],
\eeq
where $\chi_{\sigma}^{\alpha}\equiv\avg{\sum_{\p}e^{i\alpha\pi/3} e^{-i\vec{p}\cdot\vec{v}_{\alpha}} f^{\dagger}_{\tn{MM},\sigma} (\p) f_{\tau(\alpha),\sigma}(\p)}$. The total mean-field Hamiltonian is obtained by combining Eq.\ref{spinon_pp} and Eq.\ref{spinon_ph} as $H^{\tn{MF}}_{\tn{spinon}} = H^{\tn{pp}}_{\tn{chiral}} + H^{\tn{ph}}_{\tn{chiral}}$.

\subsubsection{Bosonic mean-field Hamiltonian}

\begin{figure}[pth!]
\centering
\includegraphics[width=0.5\linewidth]{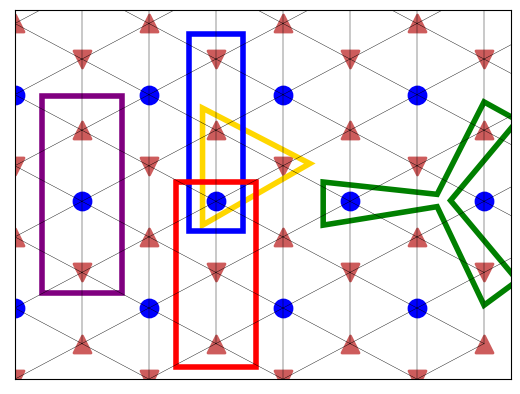} 
\refstepcounter{suppfigure}
\caption{{\bf Clusters for parton-mean field theory.} Illustration of 3-site clusters with mutually connected sites.} 
\label{fig:cluster}
\end{figure}

To deal with the quartic interaction terms of the bosonic mean-field Hamiltonian for the three-orbital model, we utlize a cluster approximation \cite{Zhao_PRB_2007}. Since we are interested in situations where the total filling of 3 orbitals is given by some fixed value (i.e. there is a global $U(1)$ conservation for rotor charges $\{n_i^\theta\}$), we go beyond the usual 2-site cluster approximation and consider 3-site clusters that comprise each of the three orbitals. To be specific, for sites $(i,j,k)$ within each cluster illustrated in Fig.~\ref{fig:cluster}, we construct a cluster Hamiltonian:
\beq 
H^\theta_{ijk} &=&-\sum_{r\in \{ijk\}}\left(\sum_{\ell\notin \{ijk\}}\sum_\sigma t_{r\ell,\sigma} \chi_{r\ell,\sigma}b^\dagger_{r}\langle b_{\ell} \rangle +\tn{h.c.} + \frac{U}{2} (2-n_r^\theta)(1-n_r^\theta)-  \mu^\theta_{ijk}n^\theta_r  +\alpha \delta_r (2-n_r^\theta) \right)   \nn \\
&+&\sum_{r\ne r'\in\{ijk\}}\left(V_{rr'}(2-n_r^\theta)(2-n_{r'}^{\theta})- \sum_\sigma t_{rr',\sigma}\chi_{rr',\sigma}b^\dagger_r b_{r'} +\tn{h.c.}\right),
\eeq 
where $\langle b_\ell\rangle$ denotes the bosonic superfluid order parameter for the site $\ell$ outside of the cluster, and $\mu^\theta_{ijk}$ for the cluster $\{ijk\}$ is chosen to impose the number constraint, $\sum_{r\in\{ijk\}}n^\theta_{r}=\tn{const}$.

\section{Layer-polarized Mott insulator at high displacement fields}
\label{sec:SI_LPMI}

\begin{figure}[pth!]
\centering
\includegraphics[width=0.4\linewidth]{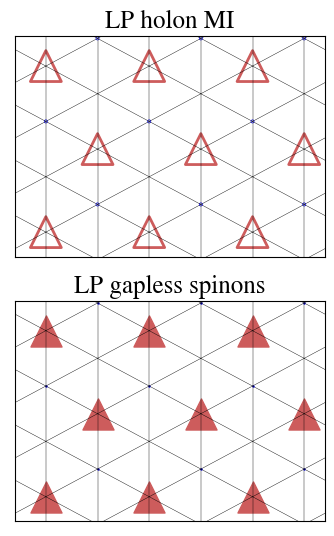} 
\refstepcounter{suppfigure}
\caption{{\bf Local holon and spinon occupations at large displacement-field.} The local occupation for the holons (upper panel) and spinons (lower panel) in the layer-polarized Mott insulator at higher fields.} 
\label{fig:LPMI_SI}
\end{figure}

Deep inside the Mott insulator at $E_z \gg E_c$, both the holons and spinons are localized only on MX sites and the system eventually reduces to a layer-polarized (LP) triangular lattice. The spinon pairing is lost due to its LH character, and the spinons are expected to form a spinon Fermi surface, if a small $t_{HH}^{(2)}$ is further applied to avoid the trivial flat-band for a large $E_z$. These trivial spinon flat-bands arise from limitations of the specific cluster approximation, and the mutual feedback between vanishing $B_{ij}$ and $\chi_{ij,\sigma}$ deep in the bosonic Mott insulating phase. In Fig.~\ref{fig:LPMI_SI}, we present the localization pattern of the holons and spinons in real-space in the LP Mott insulator phase, with the numerical data used to generate them taken from the values corresponding to the orange `$\times$' in Figs.~\ref{fig3_boson} and \ref{fig4_spinon}, respectively.

\section{Results for filling fraction away from $\nu=1$}
\label{sec:SI_nu_1+}

In this section, we briefly discuss our results obtained from parton mean-field theory for filling fraction away from $\nu=1$. As expected and discussed in the main text, superconductivity persists away from $\nu=1$, but is strongly particle-hole asymmetric. We present results for the maximal electronic pairing-channel $\Delta_{e}(\nu,E_z)$ using the full self-consistent parton mean-field theory in Fig.\ref{fig:nu_E}a. Model parameters and $\alpha_{f,\theta}$, $\beta_{f,\theta}$ are the same as Fig.\ref{fig4_spinon} in the main text.

At small displacement field, the electronic pairing gap is always finite, with the peak of the electronic pairing gap away from $\nu=1$, as shown in Fig.\ref{fig:nu_E}b. This effect can be related to the increasing superfluid density for chargon when doping from commensurate filling $\nu=1$. With increasing displacement field, a critical value of $E_z^c\sim60\tn{meV}$ shows up when the electronic pairing gap at $\nu=1$ is fully suppressed, which corresponds to the critical displacement field for the superconductor-Z$_2$ spin liquid transition. Hole doping this Z$_2$ spin liquid will restore the electronic pairing, as shown in Fig.\ref{fig:nu_E}c. Interestingly, we find that the electronic pairing is suppressed at $\nu>1$ at large $E_z$. For $\nu=1+x>0$, we find that $\nu=1$ is always favored to located at MM/MX sites to lead to a compensated excitonic order, and the rest $\nu=x$ is delocalized at XM sites. At large $E_z$, holons at MM/MX sites $(\nu=1)$ forms an excitonic Mott insulator and holons at XM sites $(\nu=x)$ forms superfluid. Thus, the electronic pairing (between MM/MX) is suppressed at large $E_z$ at $\nu>1$.

\begin{figure}[pth!]
\centering
\includegraphics[width=1.0\linewidth]{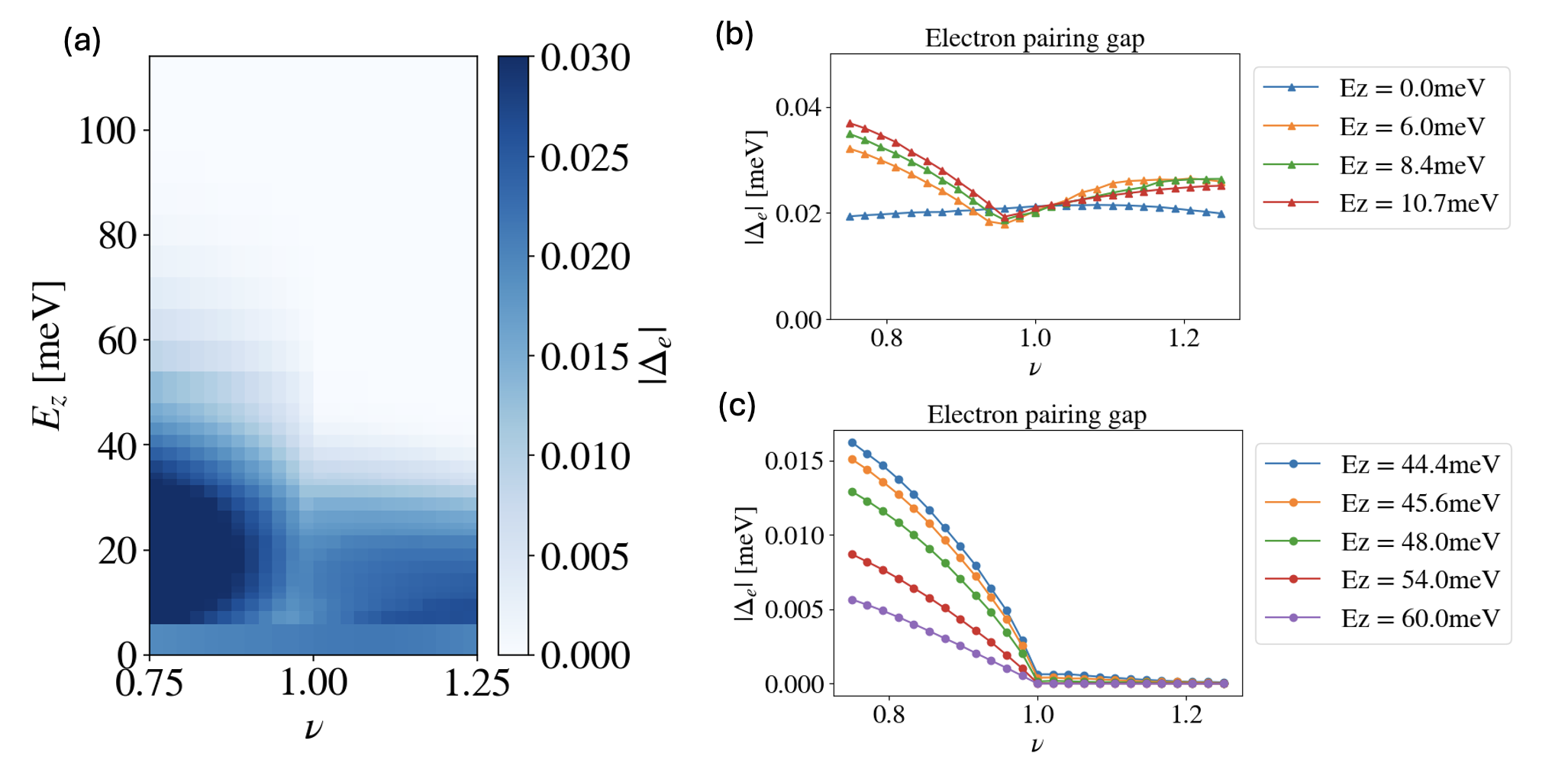} 
\refstepcounter{suppfigure}
\caption{{\bf Phase-diagram in the vicinity of $\nu=1$.} (a) Maximal electronic pairing-channel $\Delta_{e}(\nu,E_z)\equiv\tn{max}_{\alpha,S}\{ \avg{b_{\tau(\alpha)}} \avg{b_{\tn{MM}}} \Delta^{\alpha}_S \}$, obtained from full self-consistent parton mean-field computation. (b) $\Delta_{e}(\nu,E_z)$ along different cut of fixed $E_z$ in the color plot for small $E_z$. (c) $\Delta_{e}(\nu,E_z)$ along different cut of fixed $E_z$ in the color plot for large $E_z$. The model parameters used in all of our simulations are $\alpha_f/\alpha_\theta=15$ and $\beta_f/\beta_\theta=6$.}
\label{fig:nu_E}
\end{figure}

\newpage
\section{Additional results at $\nu=1$ including all three-site cluster}
\label{sec:SI_all_clusters}

In the main text, we presented the results for a single three-site cluster denoted by the yellow triangle in Fig.\ref{fig:cluster}. In this section, we show the complementary result with all three-site clusters in Fig.\ref{fig:cluster} included in the analysis of the chargon mean-field Hamiltonian. The non-universal details associated with the phase boundaries depend on the specific cluster choice for the chargon mean-field theory. However, since the relation between chargon polarizability and spinon polarizability remains unknown based on a microscopic theory, the spinon and chargon sector may respond to the displacement field and moir\'e depth differently (the parameter $\alpha_{\theta}$ and $\beta_{\theta}$ in Eq.\ref{eq:mf_tot}).

In Fig.\ref{fig:5cluster}, we present the result with all three-site-cluster taken into account in the bosonic mean-field Hamiltonian. At $E_z=0$, the system is still an extended s-wave spin-singlet superconductor. However, in contrast with the result shown in Fig.\ref{fig4_spinon}, as increasing $E_z$, the spinon pairing gap closes before the chargon enters a Mott insulating phase, which results in a renormalized Fermi-liquid regime. With a further increase in $E_z$, the chargon enters a Mott insulating phase, which leads to a U(1) QSL. Clearly, depending on the large number of (variational) parameters, various scenarios for the successive phase-transitions can be realized and future experiments with finer resolution in parameter-space will help unravel the underlying physics.

\begin{figure}[pth!]
\centering
\includegraphics[width=0.7\linewidth]{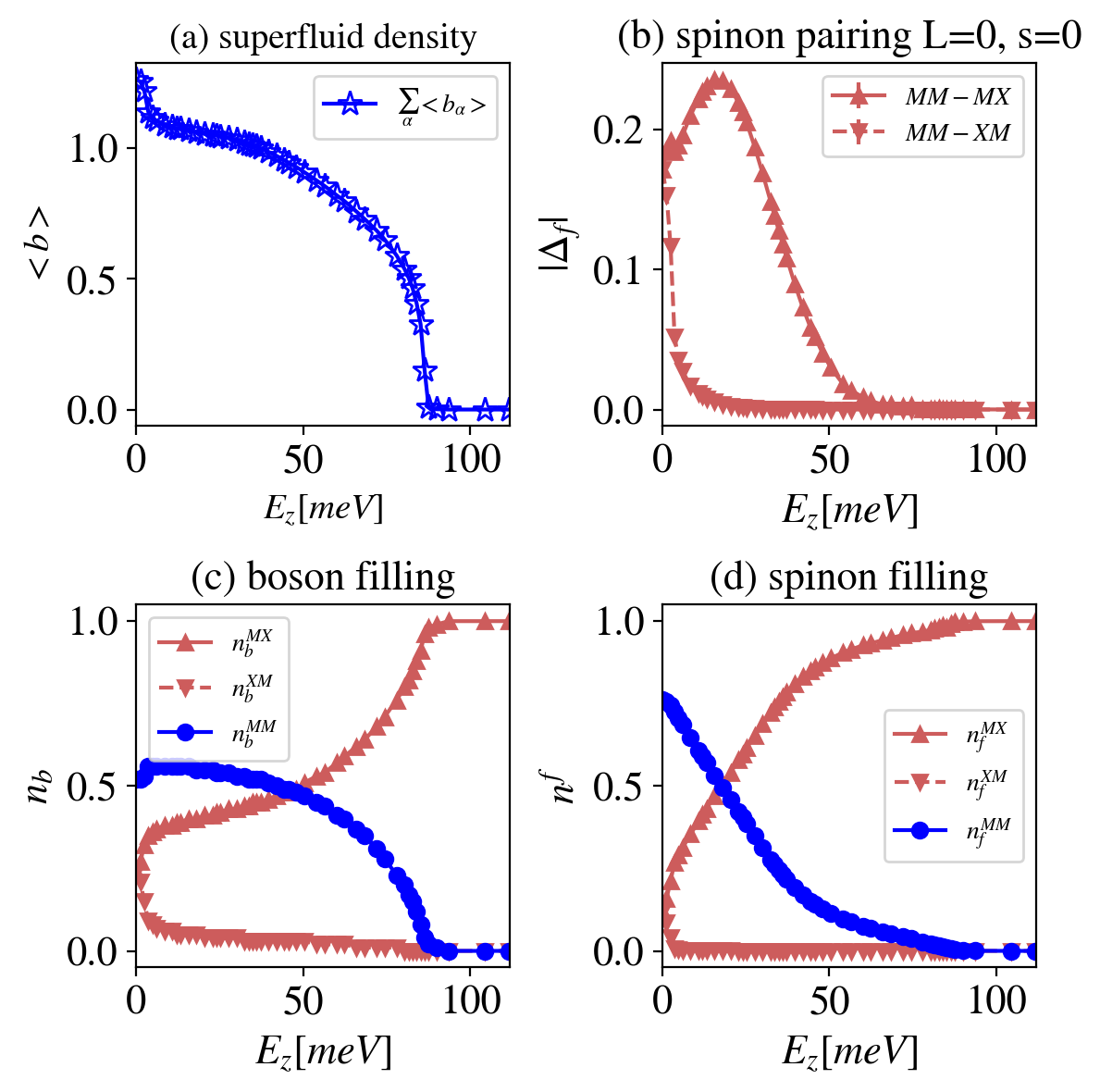} 
\refstepcounter{suppfigure}
\caption{{\bf Additional results including all three-site clusters.} Model parameters and $\alpha_{f,\theta}$, $\beta_{f,\theta}$ are the same as  Fig.\ref{fig4_spinon} in the main text. A small $t_{HH}^{(2)}\sim 1\tn{meV}$ is added to avoid the trivial flat-band in large $E_z$ case.} 
\label{fig:5cluster}
\end{figure}

\end{widetext}
\end{widetext}
\end{document}